\documentclass[showpacs,preprintnumbers,amsmath,amssymb,prb,showkeys,longbibliography]{revtex4}

\usepackage{graphicx}
\usepackage{bm}

\begin{document}

\title{Elementary excitations in undoped layered cuprates}

\author{A. V. Syromyatnikov}
\email{asyromyatnikov@yandex.ru}
\affiliation{Petersburg Nuclear Physics Institute named by B.P.Konstantinov of National Research Center "Kurchatov Institute", Gatchina 188300, Russia}

\begin{abstract}

Using the recently proposed bond-operator technique (BOT), we discuss spin dynamics of the Heisenberg spin-$\frac12$ antiferromagnet with the ring exchange and small interactions between the second- and the third-neighbor spins on the square lattice at $T=0$. This model was suggested before for description of parent compounds of high-temperature superconducting layered cuprates. BOT describes accurately short-range spin correlations in quantum systems and provides a quantitative description of elementary excitations which appear in other approaches as bound states of conventional low-energy quasiparticles. We demonstrate that besides well-known magnons (spin-1 excitations) there are three well-defined spin-0 quasiparticles in the considered model whose energies lie near the magnon spectrum. Two of them, the amplitude (Higgs) mode and the quasiparticle which we named singlon, produce pronounced anomalies observed experimentally in the Raman scattering, resonant inelastic x-ray scattering, and infrared optical absorption. We find sets of the model parameters which describe quantitatively experimental data obtained in $\rm La_2CuO_4$ and $\rm Sr_2CuO_2Cl_2$.

\end{abstract}

\pacs{75.40.Gb, 75.50.Ee, 75.10.Jm}
\keywords{Spin Dynamics; Quantum spin models; Elementary excitations}

\maketitle

\section{Introduction}

Parent compounds of high-temperature superconducting layered cuprates are Mott insulators whose magnetic properties are described by the spin-$\frac12$ antiferromagnet on the square lattice. \cite{monous} Then, investigation of the latter model is a starting point for understanding the spin dynamics which apparently plays an important role in the microscopic mechanism of the high-temperature superconductivity. \cite{monous,Keimer2015} It was suggested that the one-band Hubbard model captures the essential physics of cuprates and allows to map the electron Hamiltonian in the insulating phase to the spin Hamiltonian in which all terms are series in $t/U\alt1$, where $t$ is the electron hopping matrix element and $U$ is the on-site electron-electron repulsion (see, e.g., Refs.~\cite{anderson,auer,hubb1,cupr5,Uttt,ring1} and references therein). Apart from the largest Heisenberg exchange $J>0$ acting between nearest spins, there are smaller exchange couplings $J'>0$ and $J''>0$ between the second- and the third-neighbor spins and the four-spin ring exchange $R>0$ so that the spin-$\frac12$ Hamiltonian has the form in the leading orders in $t/U$ \cite{cupr5,Uttt}
\begin{eqnarray}
\label{ham}
&&{\cal H} = J\sum_{\langle i,j \rangle}	{\bf S}_i{\bf S}_j 
+ 
J'\sum_{\langle \langle i,j \rangle \rangle}	{\bf S}_i{\bf S}_j
+ 
J''\sum_{\langle \langle \langle i,j \rangle \rangle \rangle}	{\bf S}_i{\bf S}_j
+
R \sum_{\langle i,j,k,q \rangle}	P_{\langle i,j,k,q \rangle},\\
&&P_{\langle i,j,k,q \rangle} = 
\left( {\bf S}_i{\bf S}_j \right) \left( {\bf S}_k{\bf S}_q \right)
+
\left( {\bf S}_i{\bf S}_q \right) \left( {\bf S}_j{\bf S}_k \right)
-
\left( {\bf S}_i{\bf S}_k \right) \left( {\bf S}_j{\bf S}_q \right),
\end{eqnarray}
where ${\langle i,j,k,q \rangle}$ denote an elementary square plaquette with corners $i$, $j$, $k$, and $q$ labeled while walking around the plaquette and
\begin{eqnarray}
\label{j}
J &=& 4\frac{t^2}{U}\left( 1 - 6\frac{t^2}{U^2} \right),\\
\label{jp}
J' &=& J'' = 4\frac{t^4}{U^3},\\
\label{r}
R &=& 80\frac{t^4}{U^3}.
\end{eqnarray}
Taking into account in the Hubbard model the second- and the third-neighbor hopping parameters $t'$ and $t''$ (which are normally smaller than $t$ in cuprates) produces in spin Hamiltonian \eqref{ham} numerous (smaller) terms of the order of $1/U^3$ including rather complicated multispin interactions (see, e.g., Ref.~\cite{Uttt}). In further orders in $t/U$, corrections to parameters in Eq.~\eqref{ham} and other multispin interactions arise (see, e.g., Ref.~\cite{hubb1}). As a result, the constraint \eqref{j}--\eqref{r} on parameters is considered to be relaxed in real materials and model \eqref{ham} is often used to describe experimental data in undoped cuprates by independent varying $J$, $J'$, $J''$, and $R$. Then, it should be stressed that Eq.~\eqref{ham} is a simplified effective model of isolated cuprates which has been considered with the hope that it can catch key experimental findings.

The existence in cuprates of the noticeable ring exchange was discovered long time ago (see, e.g., Refs.~\cite{ring1,ring2}). Probably the most prominent experimental manifestation of the ring exchange is the magnon dispersion along the magnetic Brillouin zone (BZ) boundary which cannot be attributed neither to quantum fluctuations nor to $J'$ (the latter should be unrealistically large in magnitude and negative to account for this effect). \cite{cupr5,cupr3,cupr1,kataninmag} It was also proposed that the ring exchange affects considerably line shapes in the infrared spectroscopy \cite{ir2,ir3} and Raman scattering experiments \cite{katanin,raman6} in cuprates. In particular, it was obtained $R\approx0.42$ in $\rm La_2CuO_4$ and $\rm Sr_2CuO_2Cl_2$ using an analysis based on the linear spin-wave theory \cite{cupr5,cupr3,cupr1} while $R\approx0.24$ is found \cite{kataninmag} in $\rm La_2CuO_4$ using the self-consistent spin-wave calculations (unless otherwise specified, all energy values are expressed in units of $J$ below). The latter discrepancy can be attributed to pronounced quantum renormalization and to a slow convergence of $1/S$ series at $S\sim1$ when $R\sim1$ (Ref.~\cite{ringth}). 
\footnote{
For instance, it is obtained in Ref.~\cite{ringth} for the magnon energy at ${\bf k}=(\pi,0)$ and ${\bf k}=(\pi/2,\pi/2)$
$1.2+\frac{0.15}{2S}-\frac{0.19}{(2S)^2}$ 
and
$1.06-\frac{0.23}{2S}-\frac{0.57}{(2S)^2}$,
respectively, when $RS^2=0.12$, $J'=0.2$, and $J''=0$.
}

The spin dynamics in ordered phases of quantum antiferromagnets is governed by magnons (spin waves) which give sharp peaks in the inelastic neutron scattering cross section. Other excitations which can arise in antiferromagnets are commonly treated as bound states of several number of magnons. There is a number of quite sharp anomalies in infrared absorption, Raman scattering, and resonance inelastic x-ray scattering (RIXS) named "two-magnon peak", "bimagnon", and/or "continuum" which are traditionally attributed to multimagnon effects. These anomalies are not always related to any quasiparticles and are not often described quantitatively due to the lack of convenient theoretical methods for their consideration. In particular, two-magnon bound states correspond in the spin-wave theory to poles of four-particle vertexes. Then, one has to take into account an infinite number of diagrams to find these poles as series in $1/S$ that is impossible to do in most cases.

The aim of the present paper is the consideration of the "multimagnon" anomalies in experimental findings using the bond-operator theory (BOT) which is proposed recently for a quantitative discussion of dynamics in spin-$\frac12$ systems. The main advantage of the BOT is that magnons and relevant multimagnon spin-0 and spin-1 bound states are described by separate bosons that allows to find their spectra as series in $1/n$ using conventional bosonic diagrammatic technique, where $n$ is the maximum number of bosons which can occupy the (extended) unit cell. Although only $n=1$ has the physical meaning in this theory, we have shown recently that $1/n$ series converge rapidly in different spin-$\frac12$ systems which are not very close to quantum phase transition points so that first $1/n$ corrections provide at $n=1$ a quantitative agreement with experimental and numerical findings. \cite{ibot,aktersky,iboth,itri,itrih,itrij1j2} 

In Sec.~\ref{botsec}, we discuss the BOT in some detail and introduce spin correlators which are calculated below. We consider in Sec.~\ref{general} some general properties of dynamics of model \eqref{ham}. It is shown that in addition to well-known spin waves (spin-1 quasiparticles) there are three well-defined spin-0 elementary excitations whose spectra lie near magnons and who are much less known. It is demonstrated that our results on the magnon spectrum are in a good quantitative agreement with previous numerical and analytical findings. We discuss in Sec.~\ref{quant} quantities measured in experiments on Raman scattering, RIXS, and infrared optical spectroscopy. It was well established before that some of them are related to multispin dynamical correlators. We show that these correlators are expressed via Green's functions of spin-0 quasiparticles. We apply our theory to $\rm La_2CuO_4$ and $\rm Sr_2CuO_2Cl_2$, choose parameters of model \eqref{ham} for these materials, and find a good quantitative agreement with previous experimental results. In particular, we demonstrate that some anomalies in experimental data are produced by spin-0 quasiparticles whereas others originate from incoherent two-particle continua. Sec.~\ref{conc} contains a summary and a conclusion.

\section{Bond operator technique}
\label{botsec}

The main idea of the BOT is to take into account all spin degrees of freedom in the (extended) unit cell containing several spins 1/2 by building a bosonic spin representation reproducing the spin commutation algebra. A general scheme of construction of such representation for arbitrary number of spins in the unit cell is described in detail in Ref.~\cite{ibot}. We consider now briefly the main steps of this procedure for four-spin extended unit cell having the form of a plaquette which we use in the present study. Notice that we expand the magnetic unit cell twice and quadruple the crystal unit cell.

\begin{figure}
\includegraphics[scale=0.4]{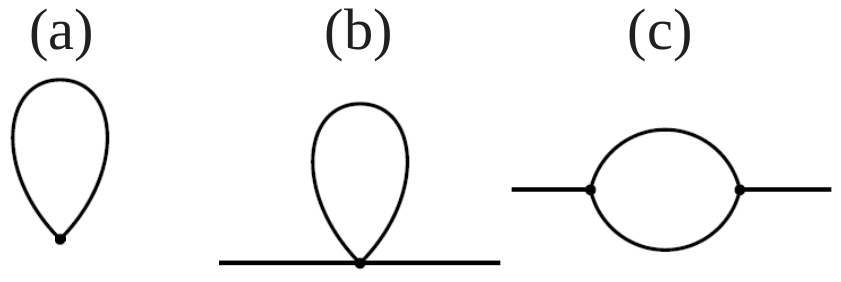}
\caption{Diagrams giving corrections of the first-order in $1/n$ to (a) the ground state energy and the staggered magnetization, and (b), (c) to self-energy parts. \label{diag}}
\end{figure}

First, we introduce 16 basis functions of four spins: $|0\rangle$ and $|a_i\rangle$ with $i=1,2,3,4,5$ from the sector with $S_z=0$, functions $|b_i\rangle$ ($|\tilde b_i\rangle$) with $i=1,2,3,4$ from the sector with $S_z=+1$ ($S_z=-1$), and states $|c\rangle$ ($|\tilde c\rangle$) from $S_z=+2$ ($S_z=-2$) sector, where $S_z$ is the projection of the total spin on a quantized axis. Then, we introduce 15 Bose operators in each unit cell which act on these basis functions according to the rule
\begin{subequations}
\label{bosons}
\begin{eqnarray}
\label{a}
a_i^\dagger |0\rangle &=& |a_i\rangle, 
\quad 
i=1,2,3,4,5,\\
\label{b}
b_i^\dagger |0\rangle &=& |b_i\rangle, 
\quad 
\tilde b_i^\dagger |0\rangle = |\tilde b_i\rangle, 
\quad 
i=1,2,3,4,\\
\label{c}
c^\dagger |0\rangle &=& |c\rangle, 
\quad
\tilde c^\dagger |0\rangle = |\tilde c\rangle.
\end{eqnarray}
\end{subequations}
where $|0\rangle$ is a selected state playing the role of the vacuum. Thus, $a$, $b$ ($\tilde b$), and $c$ ($\tilde c$) describe spin-0, spin-1, and spin-2 excitations, respectively. Then, we build the bosonic spin transformation in the unit cell as it is described in Ref.~\cite{ibot} which turns out to be quite bulky and which is presented in Ref.~\cite{ibot}. The code in the Mathematica software which generates this representation is also available in Ref.~\cite{supp2}. There is a formal artificial parameter $n$ in this spin transformation that appears in operator 
$\sqrt{n-\sum_{i=1}^5a_i^\dagger a_i-\sum_{i=1}^4(b_i^\dagger b_i-\tilde b_i^\dagger \tilde b_i) -c_ic_i^\dagger -\tilde c_i\tilde c_i^\dagger}$ 
by which linear in Bose operators terms are multiplied (cf.\ the term $\sqrt{2S-a^\dagger a}$ in the Holstein-Primakoff representation). It prevents mixing of states containing more than $n$ bosons and states with no more than $n$ bosons (then, the physical results of the BOT correspond to $n=1$). Besides, all constant terms in our spin transformation are proportional to $n$ whereas bilinear in Bose operators terms do not depend on $n$ and have the form $a_i^\dagger a_j$. We introduce also separate representations via operators \eqref{bosons} for terms ${\bf S}_i{\bf S}_j$ and $({\bf S}_i{\bf S}_j)({\bf S}_k{\bf S}_q)$ in the Hamiltonian in which $i$, $j$, $k$, and $q$ belong to the same unit cell. Constant terms in these representations are proportional to $n^2$ and terms of the form $a_i^\dagger a_j$ are proportional to $n$. \cite{ibot} Thus, we obtain a close analog of the conventional Holstein-Primakoff spin transformation which reproduces the commutation algebra of all spin operators in the unit cell for all $n>0$ and in which $n$ is the counterpart of the spin value $S$. In analogy with the spin-wave theory (SWT), expressions for observables are found in the BOT using the conventional diagrammatic technique as series in $1/n$. This is because terms in the Bose-analog of the spin Hamiltonian containing products of $p$ Bose operators are proportional to $n^{2-p/2}$ (in the SWT, such terms are proportional to $S^{2-p/2}$). For instance, to find the ground-state energy, the staggered magnetization, and self-energy parts in the first order in $1/n$ one has to calculate diagrams shown in Fig.~\ref{diag} (as in the SWT in the first order in $1/S$).

The present version of the BOT which is introduced in Ref.~\cite{ibot} contains parameters $\alpha$ and $\beta$ that allows to describe both the N\'eel ordered phase and the phase with the singlet ground state appearing when the inter-plaquette coupling is weakened (as well as the transition between these phases). In the vacuum state $|0\rangle$, parameters $\alpha$ and $\beta$ control the mixing of three plaquette states from sectors with $S_z=0$ and the total spin $S=0$, 1, and 2. As a result, the bosonic variant of the spin Hamiltonian depends on $\alpha$ and $\beta$ after the application of the spin representation. One has to calculate $\alpha$ and $\beta$ at any given model parameters by minimization of the ground state energy and find oneself as a result in one of the two phases (thus, $\alpha$ and $\beta$ appear to be functions of model parameters and series in powers of $1/n$). Then, one can come smoothly from the disordered singlet ground state to the N\'eel ordered phase upon increasing of the inter-plaquette coupling. In particular, we reproduce in this way the N\'eel ordering in the ground state of model \eqref{ham} at $J'=J''=R=0$ with the staggered magnetization $\approx0.3$ in a quantitative agreement with previous numerical and analytical findings. \cite{ibot}

Previous applications of the BOT to two-dimensional spin-$\frac12$ systems well studied before both theoretically (by other numerical and analytical methods) and experimentally show that first $1/n$ terms in most cases give the main corrections to renormalization of observables if the system is not very close to a quantum critical point. \cite{ibot,aktersky,iboth,itri,itrih,itrij1j2} Similarly, first $1/S$ corrections in the SWT frequently make the main quantum renormalization of observable quantities even at $S=1/2$ (see, e.g., Ref.~\cite{monous}). Importantly, because the spin commutation algebra is reproduced in our method at any $n>0$, $\alpha$, and $\beta$, the proper number of Goldstone excitations arises in ordered phases in any order in $1/n$ (unlike the vast majority of other versions of the BOT proposed so far \cite{ibot}).

As quantum spin correlations inside the unit cell are taken into account accurately within the BOT, we achieved in our previous works quite precise description of salient features of short-wavelength quasiparticles in different systems some of which cannot be described even qualitatively using conventional analytical approaches. \cite{iboth,itri,itrih,itrij1j2} Although the BOT is technically very similar to the SWT, the main disadvantage of this technique is that it is very bulky (e.g., the part of the Hamiltonian bilinear in Bose operators contains more than 100 terms) and it requires time-consuming numerical calculation of diagrams.

Notice that the BOT allows to consider numerous complex excitations which can arise in standard approaches as bound states of conventional quasiparticles (magnons or triplons). \cite{ibot} For instance, there are only four bosons in the four-spin BOT which describe conventional magnons. The remaining 11 bosons are responsible for some other excitations which are built on quantum states of the whole unit cell (see Eq.~\eqref{bosons}). In common methods, discussion of the bound states requires analysis of some infinite series of diagrams for vertexes. As a result, spectra of bound states cannot be obtained as series of some parameter within conventional approaches because it is normally impossible to take into account all the required diagrams. In contrast, the existence of separate bosons in BOTs (describing some bound states in considered models) allows to find their spectra as series in $1/n$ by calculating the same diagrams as for the common quasiparticles (e.g., diagrams shown in Figs.~\ref{diag}(b) and \ref{diag}(c) in the first order in $1/n$). For instance, in the ordered phase of the isotropic square-lattice antiferromagnet, the version of the BOT with two-site unit cell contains three bosons describing  two spin-1 excitations (conventional magnons) and one spin-0 quasiparticle (the Higgs mode). \cite{ibot} The price to pay for the increasing of the quasiparticles zoo in the theory is its cumbersomeness.

We take into account below diagrams shown in Fig.~\ref{diag}(b) and \ref{diag}(c) to find all self-energy parts $\Sigma(\omega,{\bf k})$ in the first order in $1/n$. We use (bare) Green's functions of the harmonic approximation in these calculations. Spectra of elementary excitations are obtained by finding zeroes of Green's functions denominators taking into account $\omega$-dependence of self-energy parts. This scheme of calculations proved to give an accurate description of all elementary excitations. \cite{iboth,itri,itrih,itrij1j2}

It should be noted that the four-site variant of the BOT used in the present study is based on the artificial enlargement of the unit cell that breaks the translational symmetry. As a result, some artifacts appear in the results obtained because it is impossible to find entire $1/n$ series for observable quantities. In particular, we find at $R=0.25J$, $J'=0.03J$, and $J''=0$ that the average value $\langle {\bf S}_i {\bf S}_j \rangle$ has the form 
$ -0.372 n^2 + 0.037 n $ and $ -0.21 n^2 - 0.108 n $ 
for the neighboring sites $i$ and $j$ inside the same plaquette and between adjacent plaquettes, respectively. These results read at $n=1$ as $-0.335$ and $-0.318$. It is the consequence of the artificial symmetry breaking that these values differ. However they differ 44\% and only 5\% in the harmonic approximation (i.e., in the leading order in $1/n$) and in the first order in $1/n$, respectively. Thus, the artificial breaking of the lattice symmetry intrinsic to the BOT does not produce significant artifacts in the first order in $1/n$ (a similar situation arises with magnon spectra which is discussed in more detail in Ref.~\cite{ibot} and below). 

\begin{figure}
\includegraphics[scale=0.1]{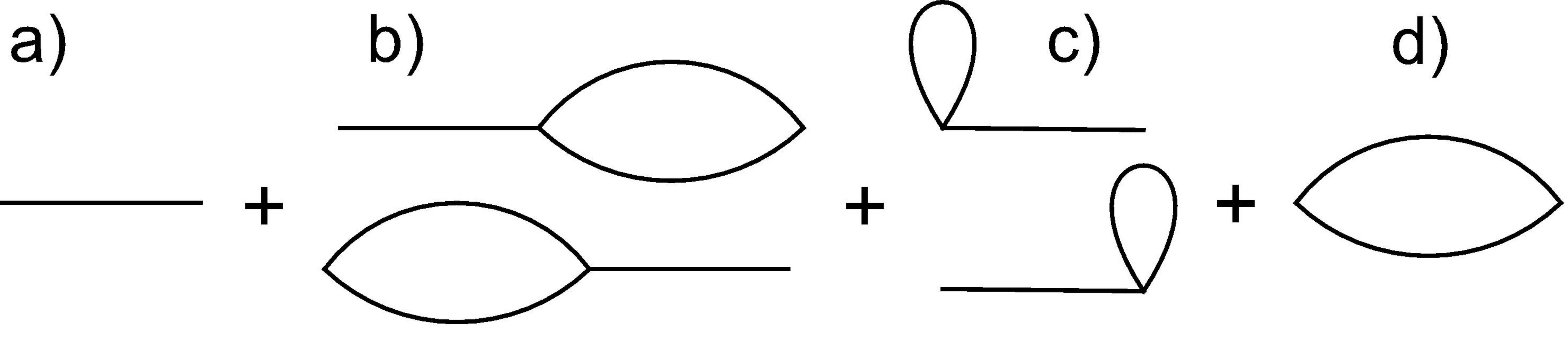}
\caption{Diagrams for dynamical spin correlators \eqref{dsf} and \eqref{dsfpp} to be taken into account in the first order in $1/n$.
\label{chifig}}
\end{figure}

We calculate below dynamical spin susceptibilities
\begin{equation}
\label{dsf}
\chi_{\alpha\beta}(\omega,{\bf k}) = 
i\int_0^\infty dt 
e^{i\omega t}	
\left\langle \left[ S^\alpha_{\bf k}(t), S^\beta_{-\bf k}(0) \right] \right\rangle,
\end{equation}
where spin operators read in our terms as 
$
S^\gamma_{\bf k} = (S^\gamma_{1\bf k} + S^\gamma_{2\bf k}e^{-ik_y/2} + S^\gamma_{3\bf k}e^{-i(k_x+k_y)/2} + S^\gamma_{4\bf k}e^{-ik_x/2})/2
$, 
the double distance between nearest neighbor spins is set to be equal to unity here and spins in the unit cell are enumerated clockwise starting from its left lower corner. We take into account diagrams shown in Fig.~\ref{chifig} in the first order in $1/n$. We will consider also some more involved dynamical correlators which are measured experimentally and which have the form
\begin{equation}
\label{dsfpp}
\chi_{PP}(\omega) = 
i\int_0^\infty dt 
e^{i\omega t}	
\left\langle \left[ P(t), P(0) \right] \right\rangle,
\end{equation}
where $P$ is a multi-spin operator. Spin-1 excitations can contribute to anomalies of $\chi_{xx}(\omega,{\bf k})$, $\chi_{yy}(\omega,{\bf k})$, and $\chi_{PP}(\omega)$ if $P$ contains an odd number of spin operators. Spin-0 quasiparticles can be studied using $\chi_{zz}(\omega,{\bf k})$ and $\chi_{PP}(\omega)$ when $P$ contains an even number of spins. We consider below in some detail $P=P^a_{\bf k}$ and $P=P^b_{\bf k}$ which are related to experimentally measured quantities, where
\begin{eqnarray}
\label{px}
P^a_{\bf k} &=& \frac{1}{\sqrt N} \sum_{\bf r} e^{-i{\bf kr}} 
{\bf S}_{\bf r} {\bf S}_{{\bf r}+{\bf a}},\\
\label{py}
P^b_{\bf k} &=& \frac{1}{\sqrt N} \sum_{\bf r} e^{-i{\bf kr}} 
{\bf S}_{\bf r} {\bf S}_{{\bf r}+{\bf b}},
\end{eqnarray}
$\bf a$ and $\bf b$ are square lattice translation vectors, and $N$ is the number of spins in the lattice. Because $P^{a,b}_{\bf k}$ contain in the leading order in $1/n$ linear combinations of some bosons $a$ (spin-0 bosons), $\chi_{PP}(\omega)$ contains at $P=P^{a,b}_{\bf k}$ Green's functions of corresponding spin-0 quasiparticles which produce visible anomalies. 

\section{Some general properties of the model dynamics}
\label{general}

\subsection{Spin-1 excitations}

\begin{figure}
\includegraphics[scale=0.4]{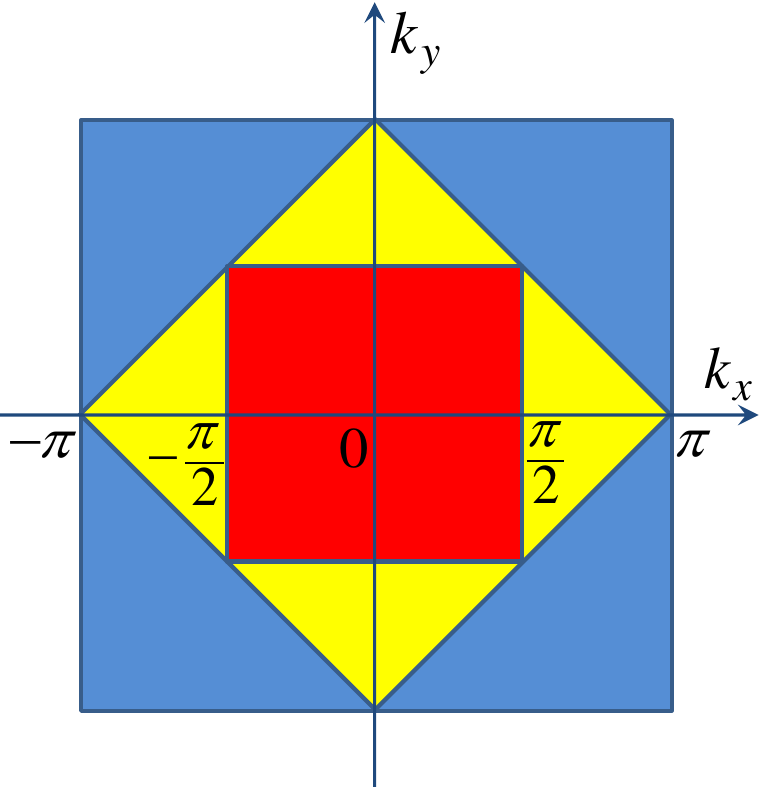}
\caption{The crystal and the magnetic Brillouin zones (BZs) are presented (the largest and the middle squares, respectively) for the simple square lattice. The distance between nearest lattice sites is set to be equal to unity. The smallest (red) square and the yellow area are the first and the second BZs, correspondingly, in the case of four sites in the unit cell.
\label{bz}}
\end{figure}

It is well known from, e.g., the spin-wave theory that there is one doubly degenerate magnon mode in the magnetic BZ shown in Fig.~\ref{bz}. The double degeneracy stems from the equivalence of excitations from sectors with $S_z=+1$ and $S_z=-1$, where $S_z$ is the projection of the total spin on the quantized axis. As it is noted above and as is discussed in detail in Ref.~\cite{ibot}, there are four bosons describing spin-1 quasiparticles in the four-spin BOT which correspond to conventional magnons so that operators $S^x_{\bf k}$ and $S^y_{\bf k}$ are linear combinations of them in the leading order in $1/n$ (and $\chi_{xx}(\omega,{\bf k})$ and $\chi_{yy}(\omega,{\bf k})$ are linear combinations of Green's functions of these bosons). To be precise, there are two couples of bosons (each couple contains one boson from sectors with $S_z=+1$ and $S_z=-1$) and spectra are equivalent of bosons from each couple. It can be shown \cite{ibot} that residues of bosons from one couple in $\chi_{xx}(\omega,{\bf k})$ and $\chi_{yy}(\omega,{\bf k})$ are zero inside the red region in Fig.~\ref{bz} and they are finite inside the yellow region whereas the situation is the opposite with the other couple of spin-1 bosons. Then, these two doubly degenerate spin-1 branches reproduce the conventional magnon spectrum in the magnetic BZ after taking into account their spectral weights in spin susceptibilities (see also Ref.~\cite{ibot} for extra detail). It should be noted however that these two branches do not completely match at the border of the red region in Fig.~\ref{bz} in the first few orders in $1/n$ (see Ref.~\cite{ibot} and below). This small mismatch is an artifact of the truncation of $1/n$ series.

\begin{figure}
\includegraphics[scale=0.38]{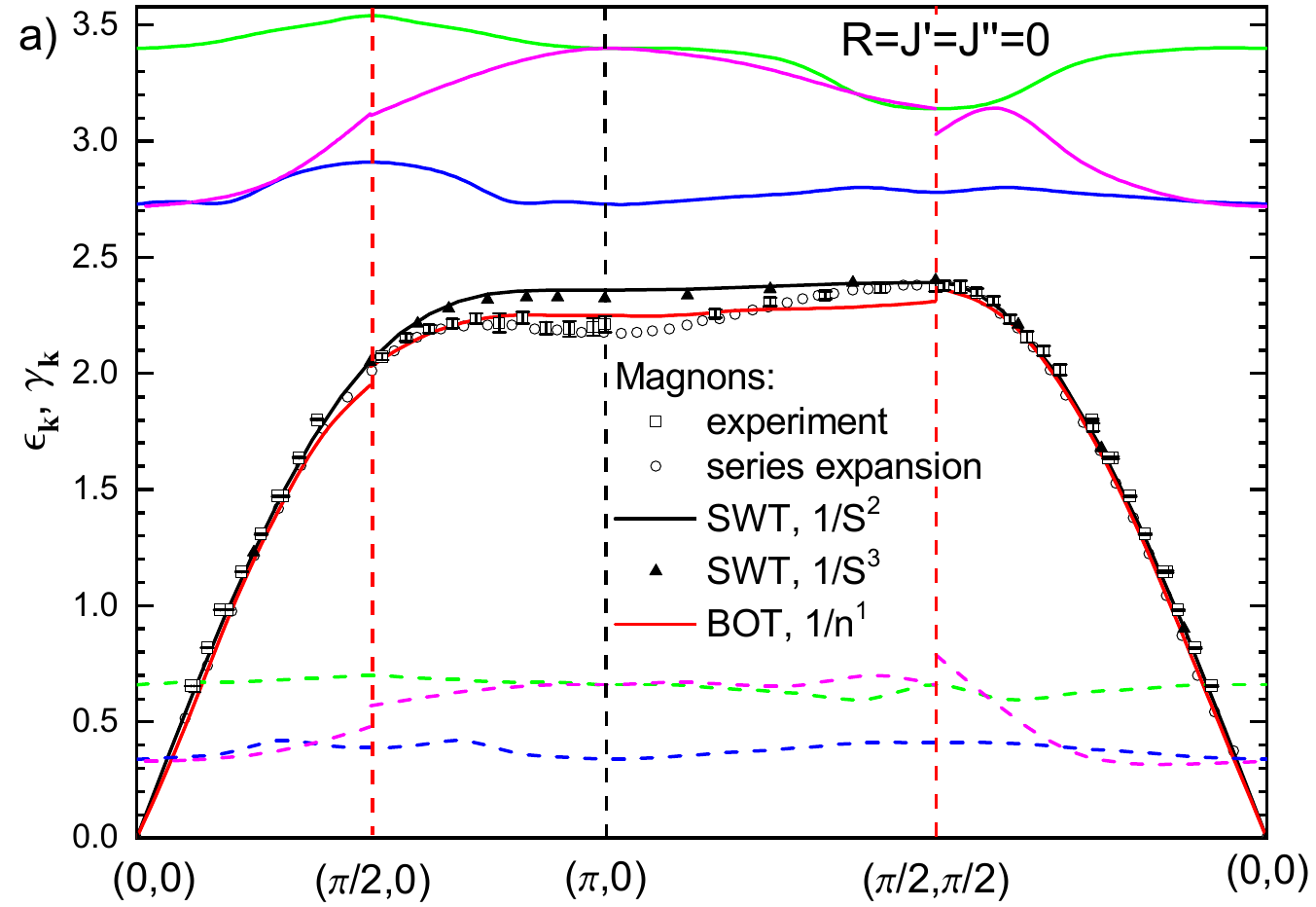}
\includegraphics[scale=0.38]{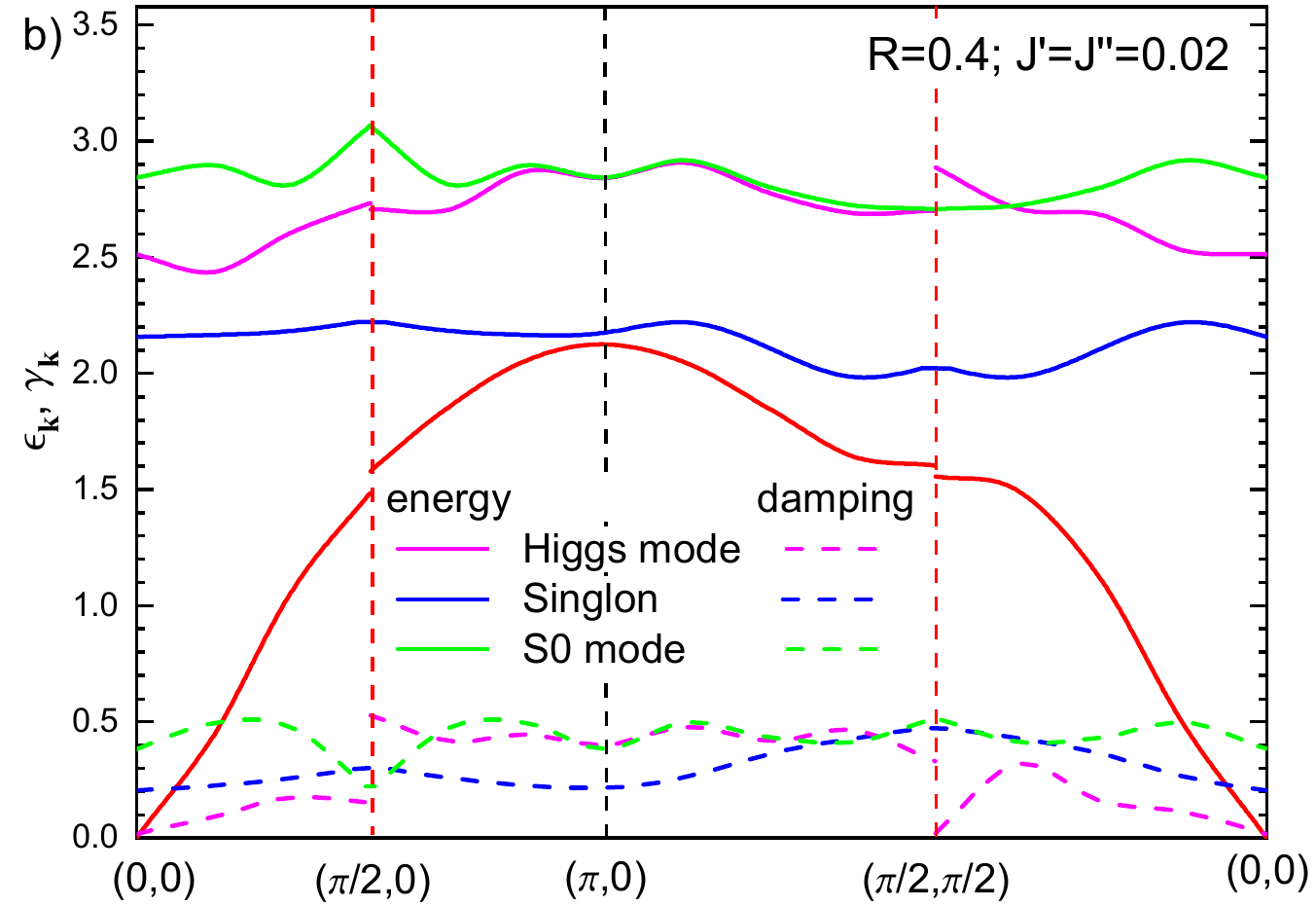}
\caption{Evolution of spectra $\epsilon_{\bf k}$ of lower energy quasiparticles in model \eqref{ham} upon varying $R$ at $J'=J''=R/20$ found using the BOT in the first order in $1/n$. The singlon, the amplitude (Higgs), and the S0 modes are well-defined spin-0 quasiparticles whose damping $\gamma_{\bf k}$ are presented with dashed lines of corresponding colors. Also shown in panel a) are magnon spectra obtained by the series expansion around the Ising limit \cite{series}, within the spin-wave theory (SWT) in the second \cite{igar,igar2} and in the third \cite{syromyat} orders in $1/S$, and neutron scattering experiment in CFTD \cite{chris1,piazza}. Borders of the first BZ with four spins in the unit cell are shown by red vertical lines (see Fig.~\ref{bz}). Breaks on these lines of spectra of magnons and the Higgs mode are discussed in the text.
\label{spectra}}
\end{figure}

We show in Fig.~\ref{spectra} the evolution of low-energy quasiparticles spectra found in model \eqref{ham} in the first order in $1/n$ at different $R$ and $J'=J''=R/20$ (see Eqs.~\eqref{jp} and \eqref{r}). As it was reported before \cite{ibot}, our results at $R=0$ for the magnon spectrum agree well with previous analytical \cite{igar,igar2,syromyat}, numerical \cite{series}, and experimental findings in the molecular magnet abbreviated as CFTD \cite{chris1,piazza}.

\begin{figure}
\includegraphics[scale=0.4]{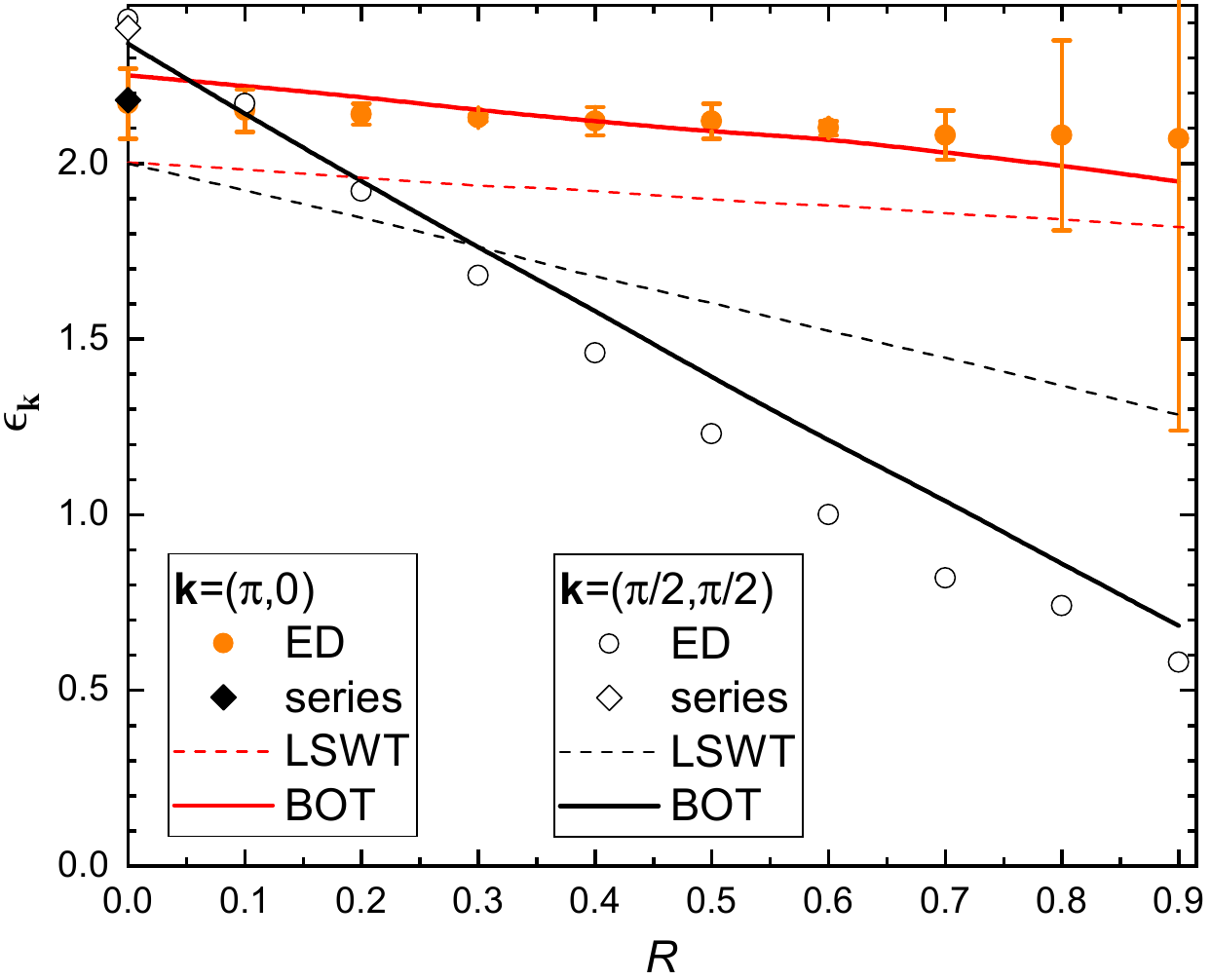}
\caption{Magnon energies at ${\bf k}=(\pi,0)$ and ${\bf k}=(\pi/2,\pi/2)$ at different values of $R$ and $J'=J''=R/20$ (see Eqs.~\eqref{jp} and \eqref{r}) found using the BOT (present study), the linear spin-wave theory(LSWT), and exact diagonalization (ED) of finite clusters containing $N\le32$ sites with the subsequent linear in $1/N$ extrapolation to thermodynamic limit (Ref.~\cite{ringnum1}). The accuracy of ED results at ${\bf k}=(\pi/2,\pi/2)$ is not estimated because data for only $N=16$ and $N=32$ are available for this momentum. \cite{ringnum1} The accuracy of ED results degrades quickly upon approaching the quantum critical point at $R\sim1$ as it is seen from the data for ${\bf k}=(\pi,0)$. Numerical results obtained at $R=0$ using series expansion around the Ising limit \cite{series} are also shown (series).
\label{mag2}}
\end{figure}

In agreement with previous spin-wave \cite{ringsw1,ringth} and numerical \cite{ringnum1} analysis, our data demonstrate the decreasing of the magnon spectrum at ${\bf k}=(\pi/2,\pi/2)$ upon $R$ increasing and a "condensation" of magnons at $R\approx 1$ signifying a phase transition. Fig.~\ref{mag2} shows a good agreement between our results for the magnon spectrum at ${\bf k}=(\pi,0)$ and ${\bf k}=(\pi/2,\pi/2)$ with previous results of the exact diagonalization of finite clusters \cite{ringnum1}. The magnon energy at momentum ${\bf k}=(\pi/2,\pi/2)$ lying on the border of the red area (see Fig.~\ref{bz}) is found in the BOT as a half-sum of two branches attributed to magnons as it is described above. Although $R$ is about a few tenths in the majority of cuprates, it has been proposed recently that the N\'eel ordered $\rm CaCuO_2$ has $R \approx 1$ (see, e.g., Ref.~\cite{rixs6} and references therein).

We do not present here spectra of other spin-1 bosons which, as it turned out, lie above spin-0 quasiparticles, have quite large damping and do not produce visible anomalies in experimental data.

\subsection{Spin-0 excitations}

We supplement previous considerations of model \eqref{ham} with the finding that there are three well-defined spin-0 quasiparticles who lie near magnons and whose spectra are also shown in Fig.~\ref{spectra}. Their spectra goes down as $R$ increases and the spectrum of one of them, singlon, appears even below magnons at some parts of the BZ at $R\agt0.5$ (we named the latter quasiparticle singlon in Ref.~\cite{ibot} because it corresponds in the harmonic approximation of the BOT to a singlet state of the plaquette propagating along the lattice). Our preliminary consideration shows that the "condensation" of magnons with momentum ${\bf k}=(\pi/2,\pi/2)$ at $R\approx 1$ is accompanied with a "condensation" of the spin-0 quasiparticle shown in green in Fig.~\ref{spectra} (S0 mode). A more detailed consideration of the transition at $R\approx 1$ is out of the scope of the present paper.

To be precise, there are four low-lying spin-0 bosons two of which we identify as two parts of the amplitude (Higgs) mode as a result of the following consideration. In the leading order in $1/n$, $S^z_{\bf k}$ is a linear combination of all spin-0 bosons except for the singlon. We show in Ref.~\cite{ibot} that the spectrum of one of this boson merges the magnon branch upon decreasing of the inter-plaquette spin interactions. As a result of this merging a gapped triplon branch appears in the magnetically disordered state having the singlet ground state. Then, this spin-0 boson describes the amplitude mode. \cite{higgs1} However its spectral weight in $\chi_{zz}(\omega,{\bf k})$ is zero inside the yellow regions shown in Fig.~\ref{bz}. There is only one another spin-0 boson whose spectral weight in $\chi_{zz}(\omega,{\bf k})$ is zero inside the red region and finite in the yellow one (the remaining spin-0 bosons appear in $\chi_{zz}(\omega,{\bf k})$ in all parts of the BZ). Then, we conclude that these two bosons are two parts of the amplitude mode with the mismatch at the border of the red area in Fig.~\ref{bz} (see Fig.~\ref{spectra}) as it was in the case of magnons considered above. As a result, the doubly degenerate magnon (Goldstone) branch and one amplitude mode arise within the BOT in the magnetic Brillouin zone as it must be according to general symmetry arguments \cite{higgs1}.

\begin{figure}
\includegraphics[scale=0.4]{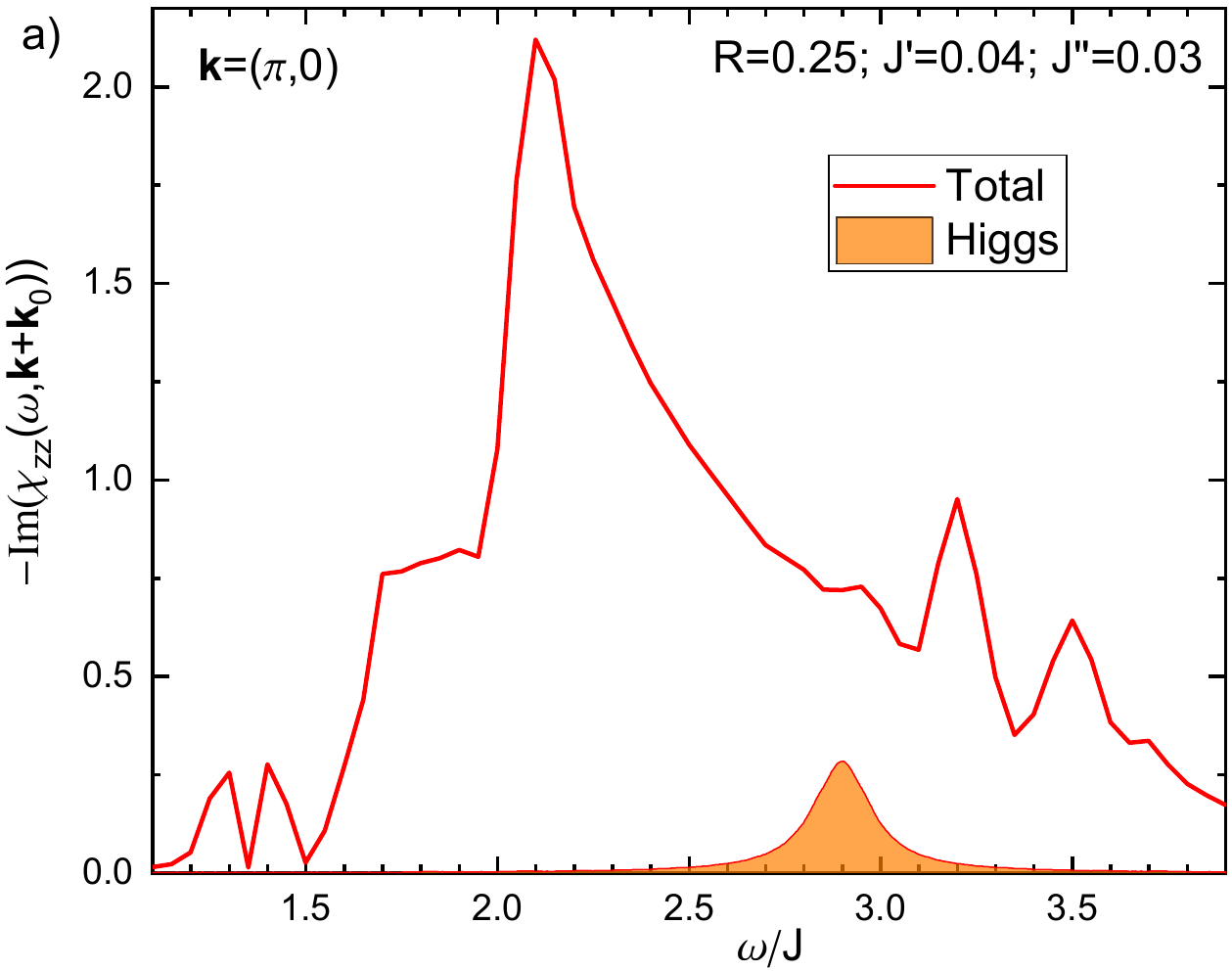}
\includegraphics[scale=0.4]{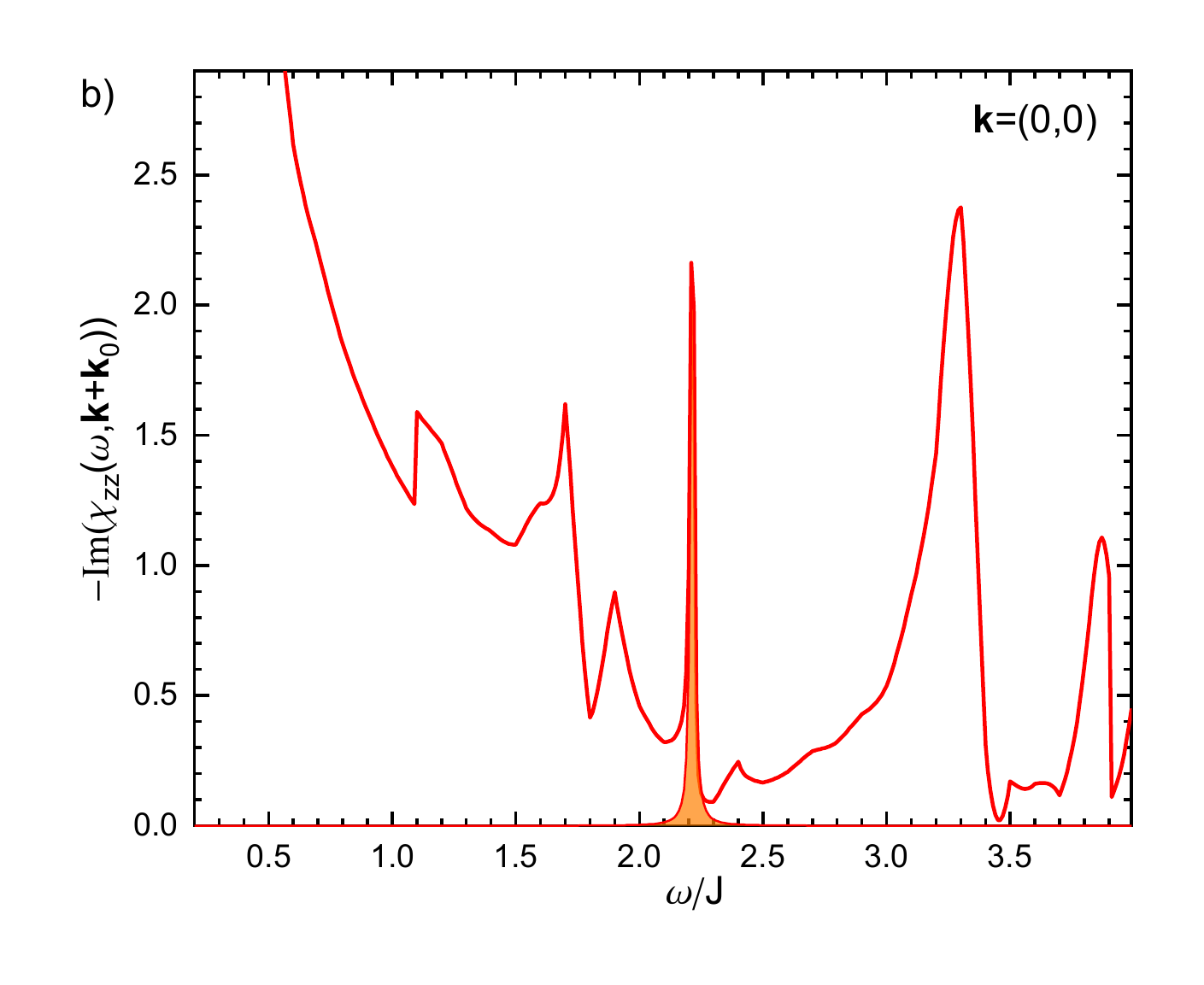}
\caption{Longitudinal dynamical structure factor (DSF) $-{\rm Im}\chi_{zz}(\omega,{\bf k})$ built on Eq.~\eqref{dsf} and found in the BOT at $R=0.25$, $J'=0.04$, $J''=0.03$, and momenta ${\bf k}+{\bf k}_0$, where ${\bf k}_0=(\pi,\pi)$. The incoherent two-particle continuum dominates in this DSF which is produced by diagrams shown in Figs.~\ref{chifig}(b) and \ref{chifig}(d). Contribution of the amplitude mode is shown which is hardly visible against the continuum.
\label{zzfig}}
\end{figure}

Although $\chi_{zz}(\omega,{\bf k})$ contains Green's functions of the Higgs and S0 modes, they are hardly visible in the dynamical structure factor (DSF) $-{\rm Im}\chi_{zz}(\omega,{\bf k})$ against the incoherent two-particle continuum produced by diagrams shown in Figs.~\ref{chifig}(b) and \ref{chifig}(d). Fig.~\ref{zzfig} illustrates this our finding. Then, it is hardly possible to study the Higgs mode in experiments with polarized neutrons which can measure the longitudinal DSF. As it is shown in Sec.~\ref{quant}, the contribution of the Higgs mode to results obtained by other experimental techniques can be more pronounced.

\begin{figure}
\includegraphics[scale=0.22]{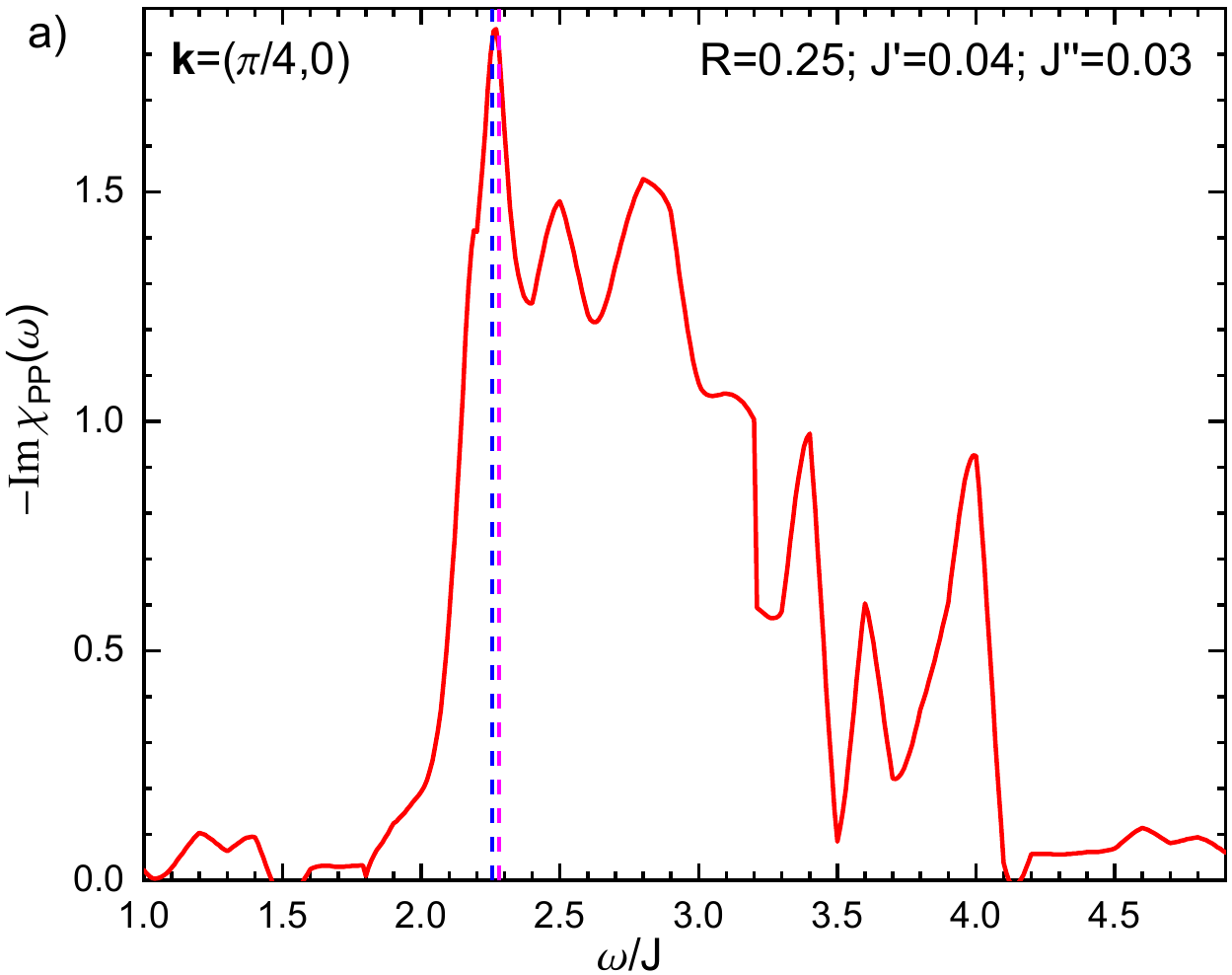}
\includegraphics[scale=0.22]{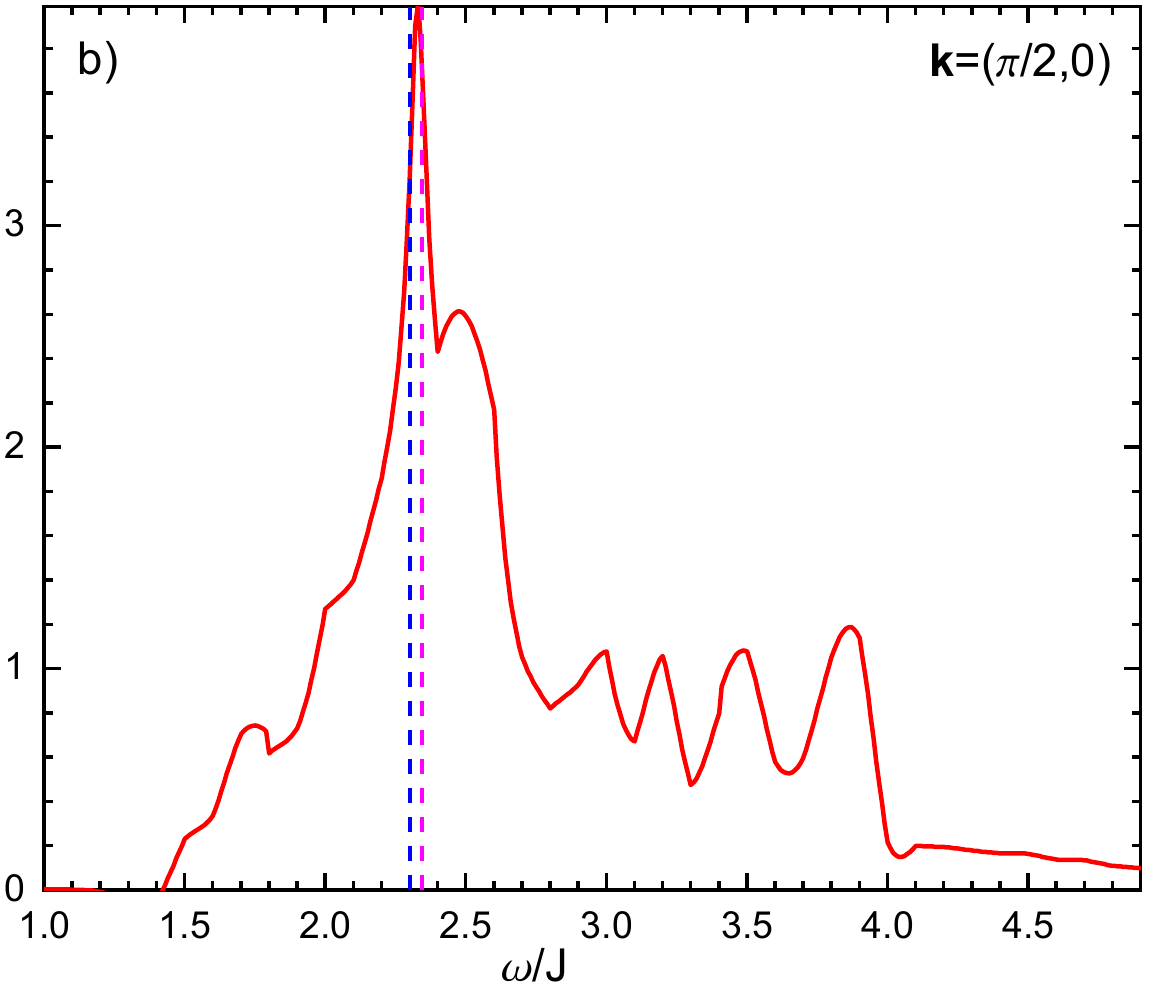}
\includegraphics[scale=0.22]{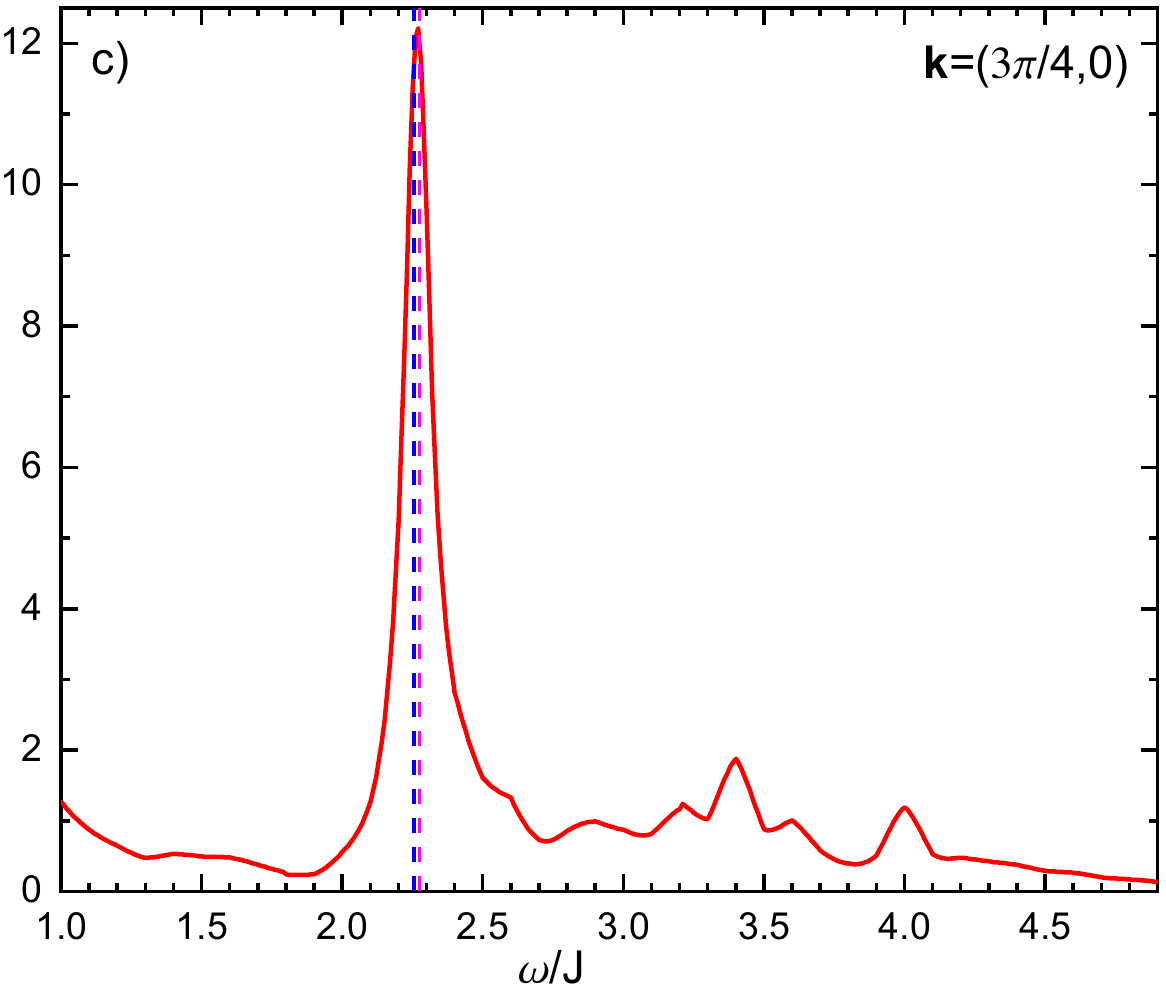}
\includegraphics[scale=0.22]{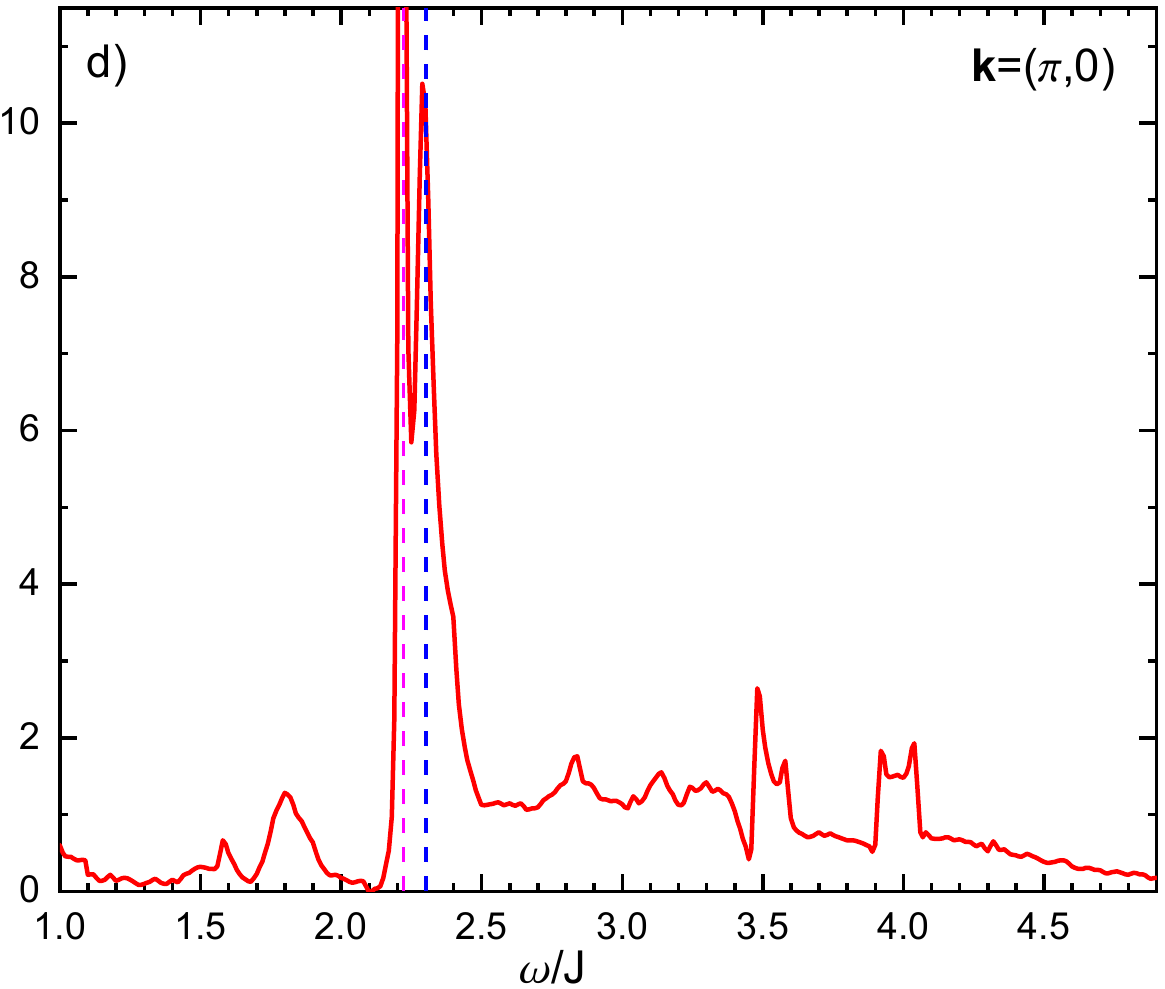}
\includegraphics[scale=0.22]{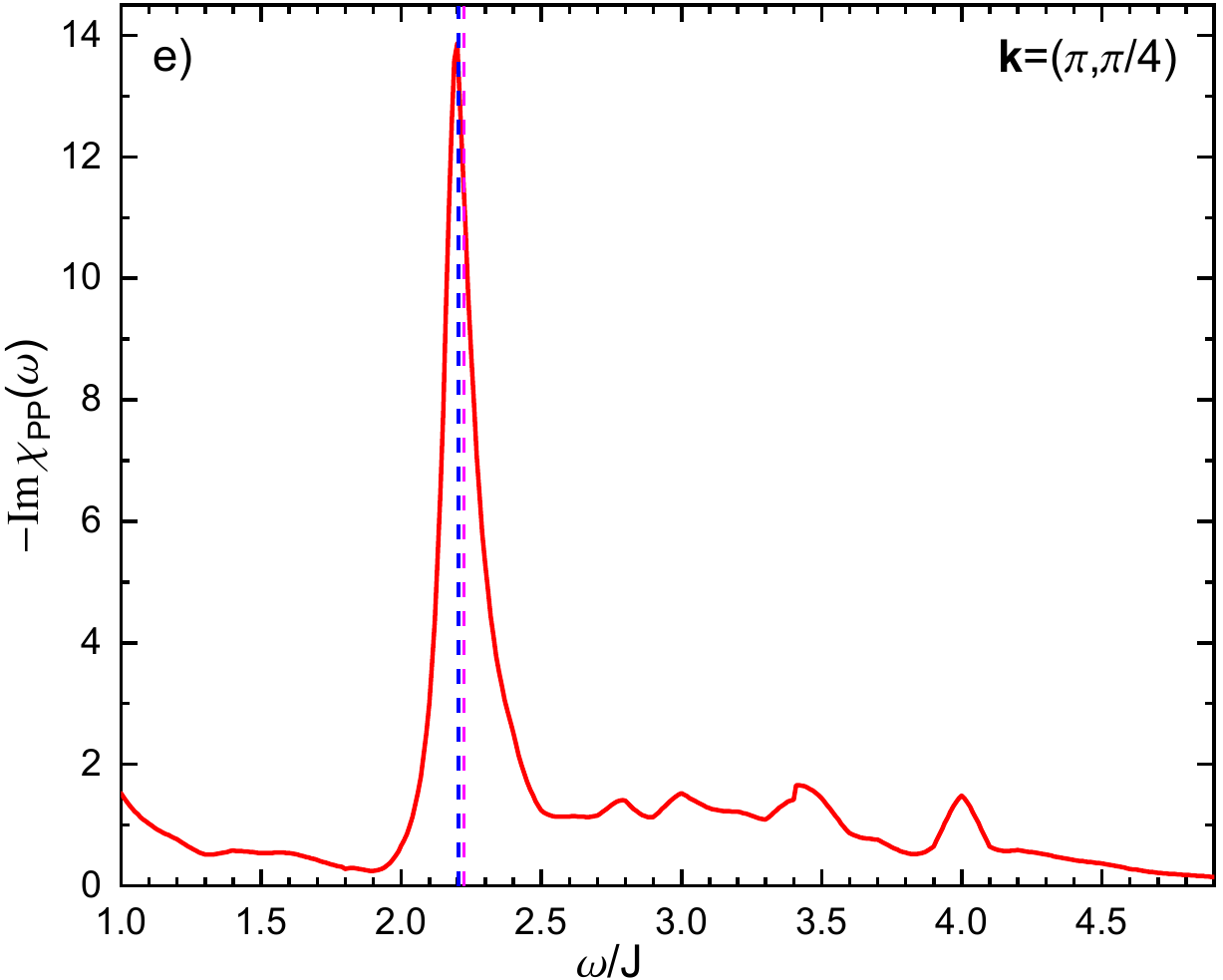}
\includegraphics[scale=0.22]{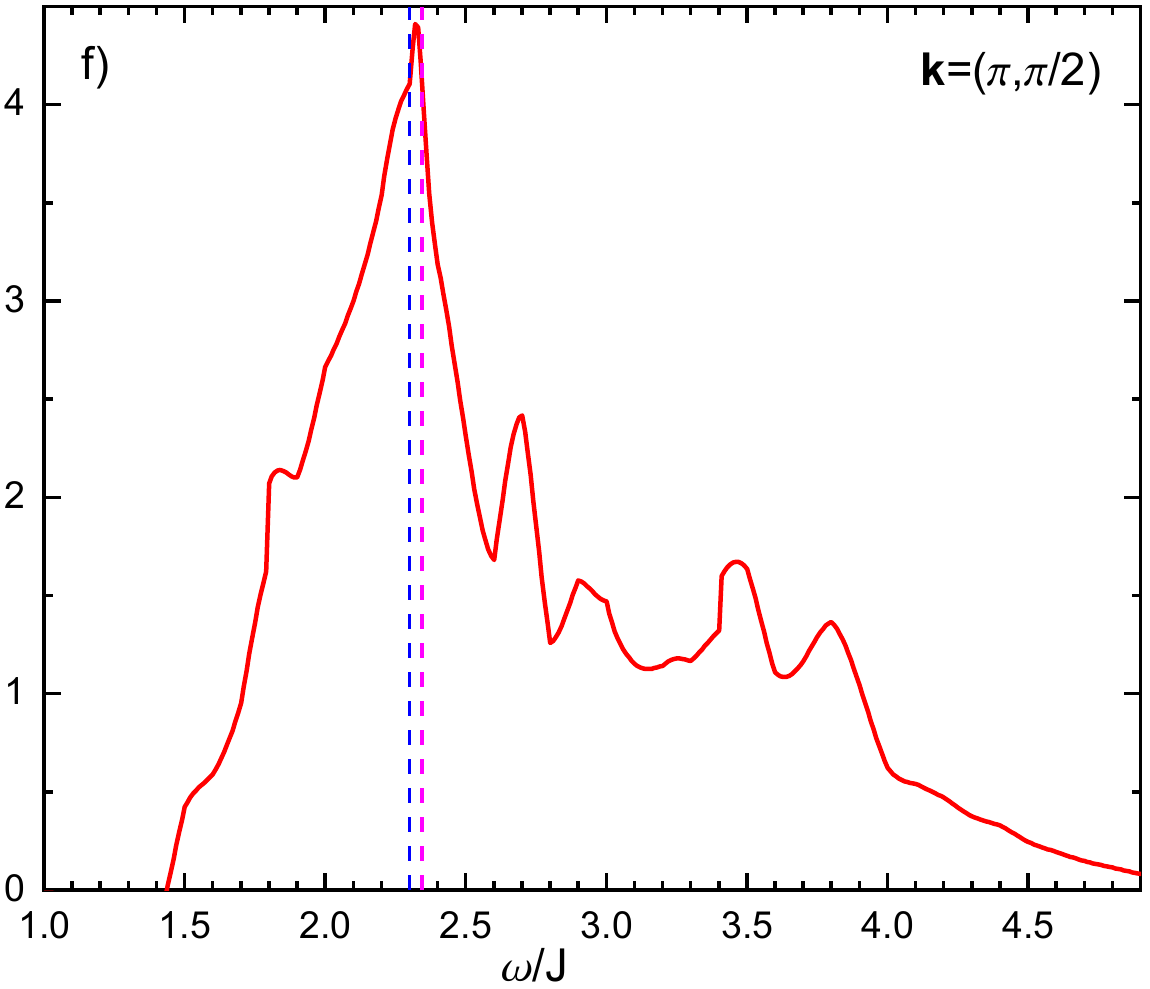}
\includegraphics[scale=0.22]{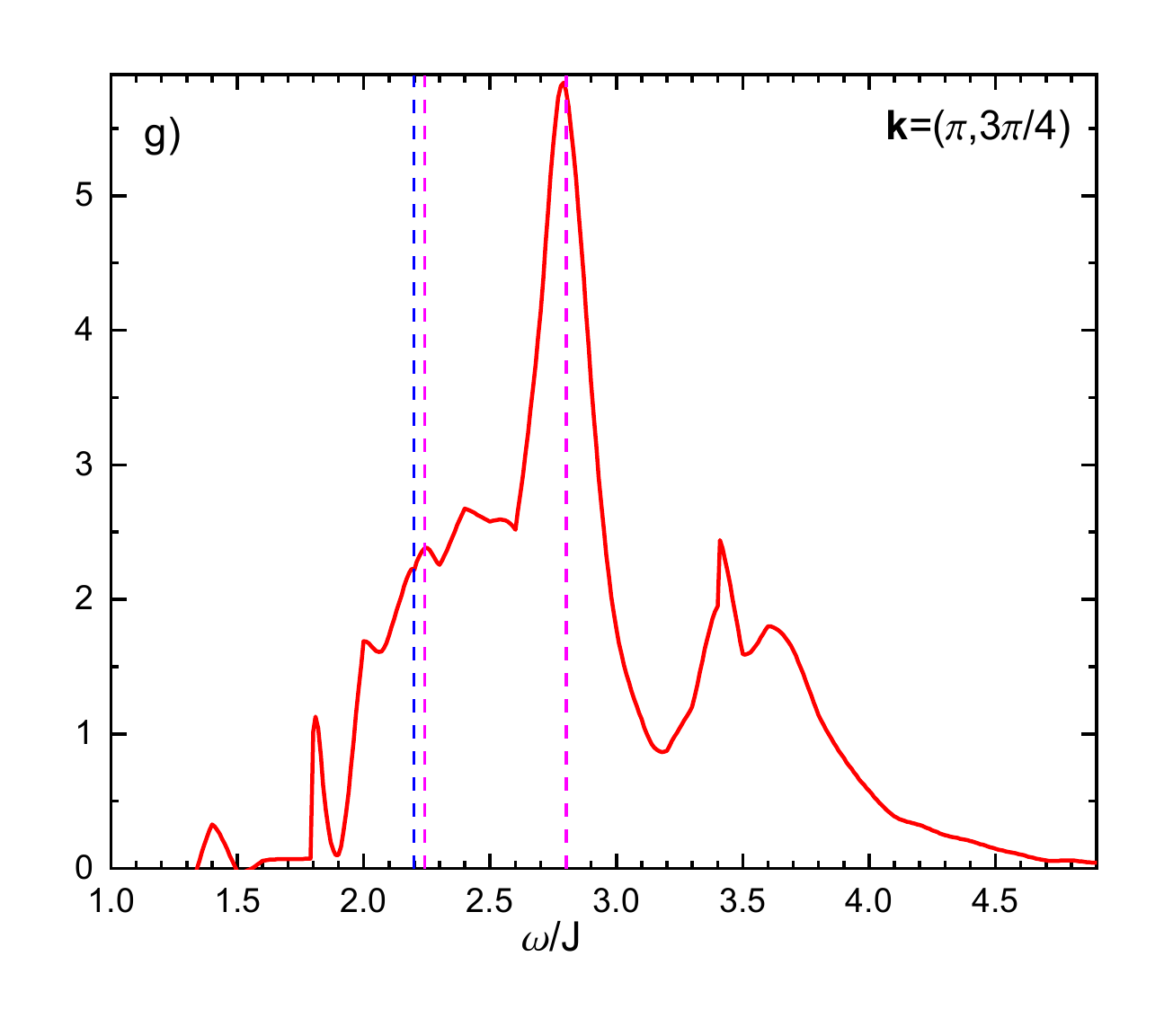}
\includegraphics[scale=0.22]{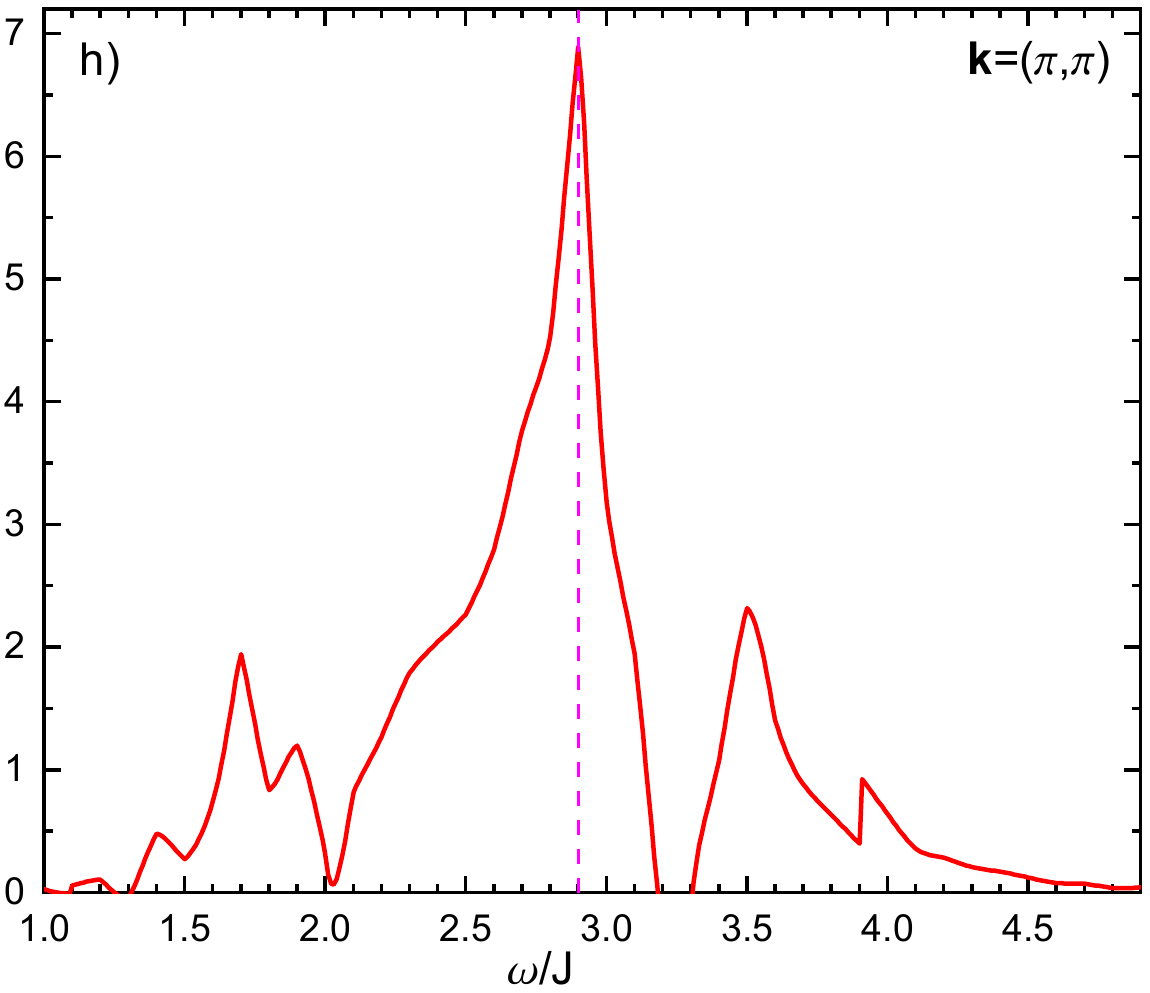}
\includegraphics[scale=0.22]{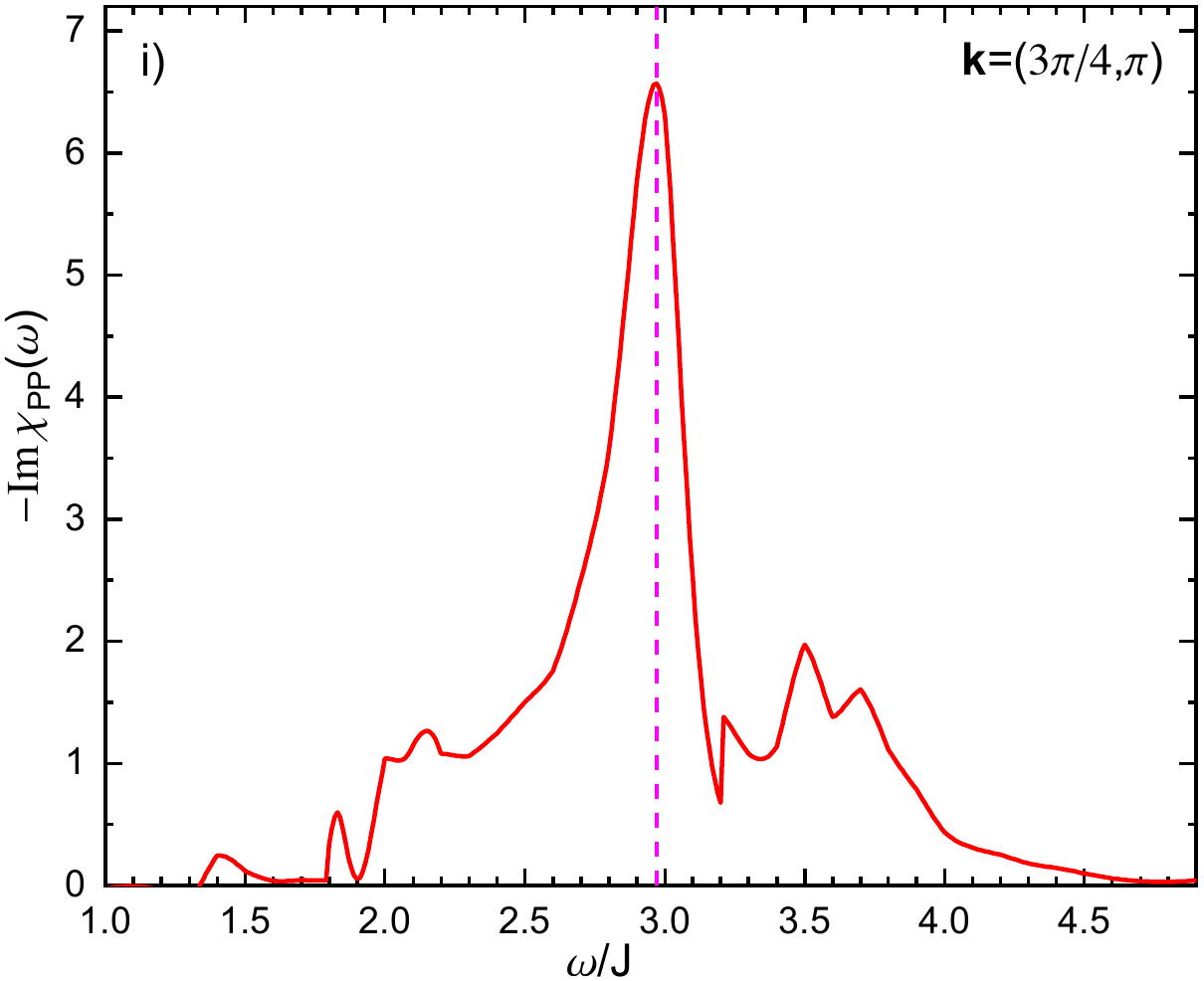}
\includegraphics[scale=0.22]{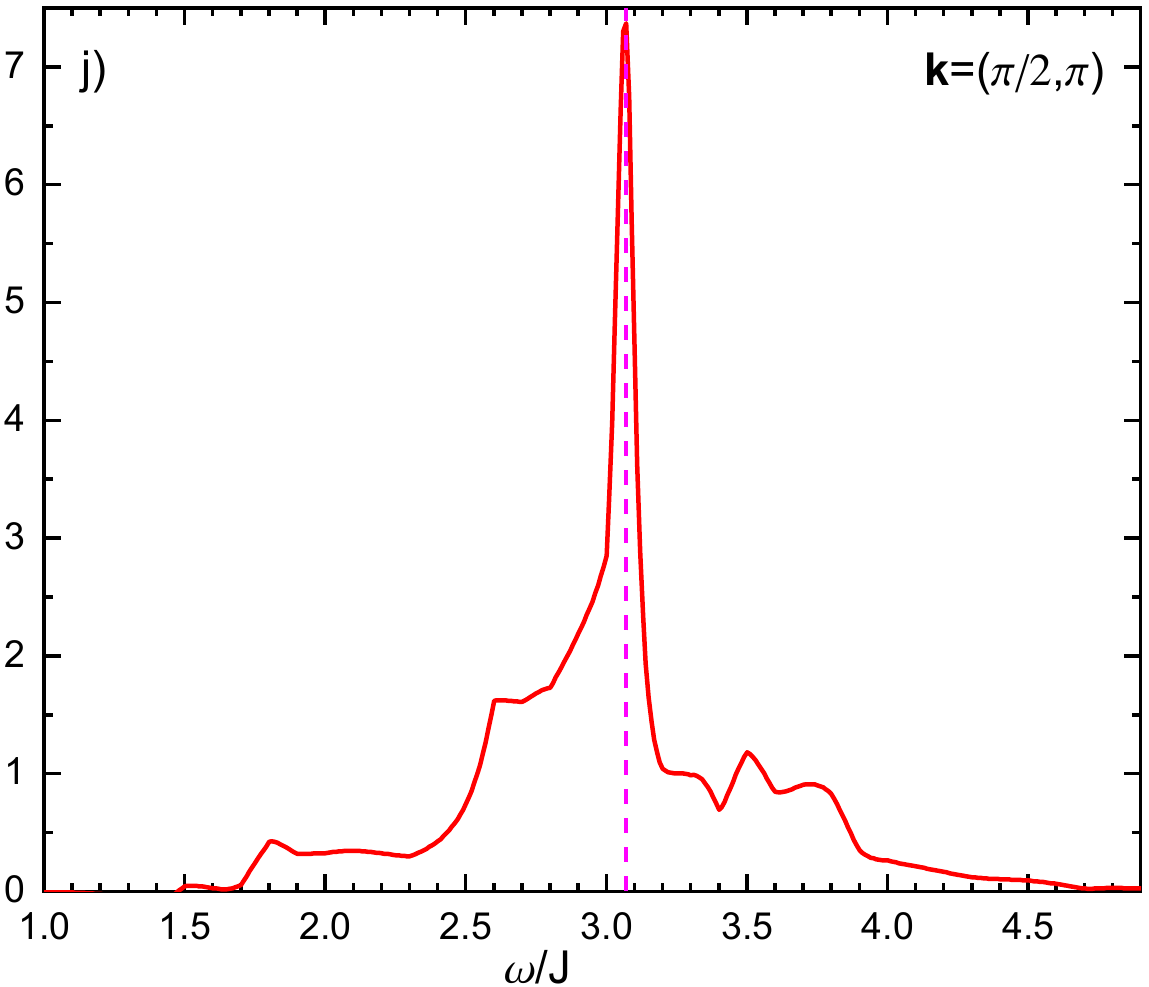}
\includegraphics[scale=0.22]{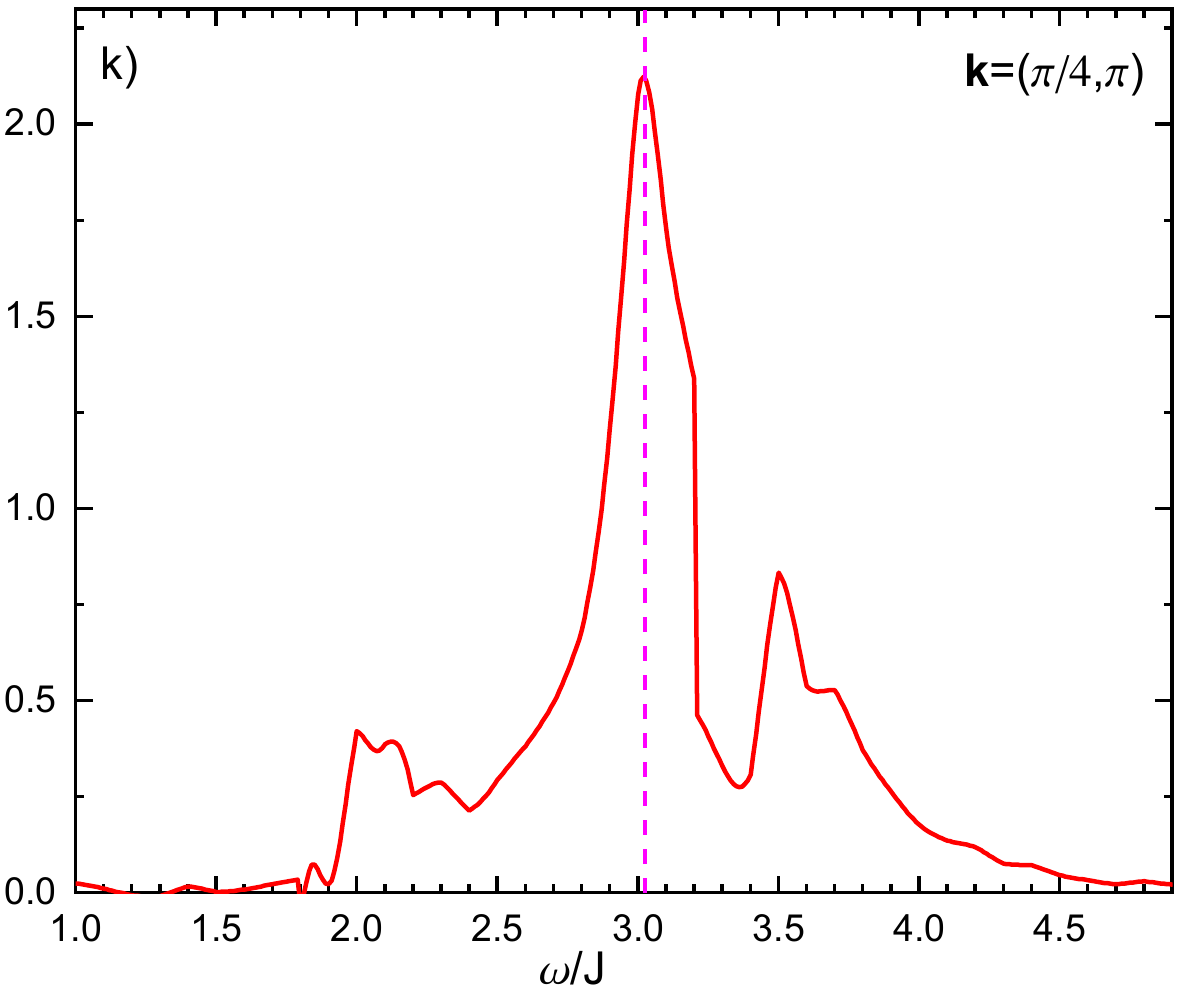}
\includegraphics[scale=0.22]{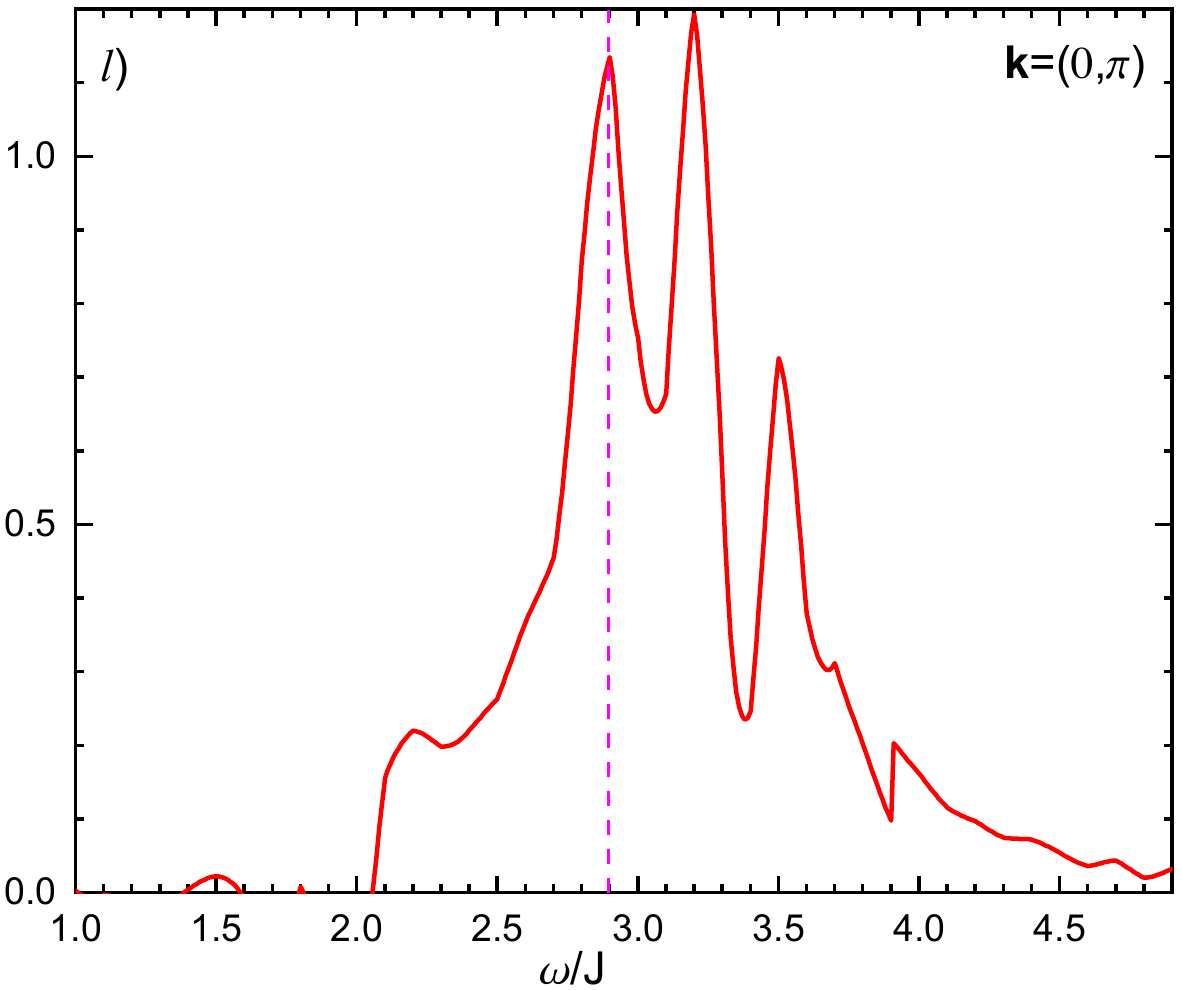}
\caption{Dynamical multispin structure factor $-{\rm Im}\chi_{PP}(\omega)$ built on Eq.~\eqref{dsfpp} for $P=P^a_{\bf k}$ (see Eq.~\eqref{px}) and found in the BOT for some momenta at $R=0.25$, $J'=0.04$, and $J''=0.03$. Positions of anomalies corresponding to quasiparticles whose spectra are shown in Fig.~\ref{spectra} are depicted by vertical dashed lines of respective colors. This quantity is measured in RIXS experiment.
\label{ppfigs}}
\end{figure}

The situation is more favorable with experimental techniques probing four-spin correlators. We plot in Fig.~\ref{ppfigs} the DSF $-{\rm Im}\chi_{PP}(\omega)$ at $P=P^a_{\bf k}$ (see Eqs.~\eqref{dsfpp} and \eqref{px}) for some momenta. It is seen that the singlon and the Higgs mode produce distinct anomalies which can be easily seen against the two-particle continuum. We demonstrate in the next section that these anomalies are really observed experimentally. The asymmetry should be noted of the considered DSF for ${\bf k}=(p,q)$ and ${\bf k}=(q,p)$ which is seen in Fig.~\ref{ppfigs} (and will be seen in experimental data discussed below). This is due to the fact that $P^a_{\bf k}$ is built on bond spin operators elongated along $\bf a$ lattice direction.

To conclude this general discussion of spin-0 elementary excitations, it would be instructive to present the following most simple operators $P_{\bf k}$ whose representations contain linear in Bose-operators terms which correspond to one sort of considered spin-0 excitations:
\begin{eqnarray}
\label{phiggs}
	P_{\bf k}^{Higgs} &=& \frac{1}{\sqrt N} \sum_{\bf r} e^{-i{\bf kr}}
	\left( S^z_{2\bf r} + S^z_{2{\bf r}+{\bf a}+{\bf b}} \right),\\
\label{ps0}
	P_{\bf k}^{S0} &=& \frac{1}{\sqrt N} \sum_{\bf r} e^{-i{\bf kr}}
	\left( S^z_{2\bf r} - S^z_{2{\bf r}+{\bf a}+{\bf b}} \right),\\
\label{psinglon}
	P_{\bf k}^{singlon} &=& P^a_{\bf k} - P^b_{\bf k},
\end{eqnarray}
where $P^{a,b}_{\bf k}$ are given by Eqs.~\eqref{px} and \eqref{py}. Susceptibilities \eqref{dsfpp} built on operators \eqref{phiggs}--\eqref{psinglon} contain Green's functions which lead to anomalies produced by only corresponding quasiparticles (in the next orders in $1/n$, diagrams shown in Figs.~\ref{chifig}(b)--\ref{chifig}(d) arise). Notice also that $P^a_{\bf k} + P^b_{\bf k}$ is another operator in which linear terms contain only the Higgs Bose-operators. It has somewhat more complex form than Eq.~\eqref{phiggs} but the anomaly produced by the Higgs quasiparticle is visible in the corresponding susceptibility \eqref{dsfpp} much better against the incoherent background.

\section{Spin excitations probed experimentally in undoped layered cuprates}
\label{quant}

\subsection{Neutron scattering. Magnons and the amplitude mode.}

\begin{figure}
\includegraphics[scale=0.4]{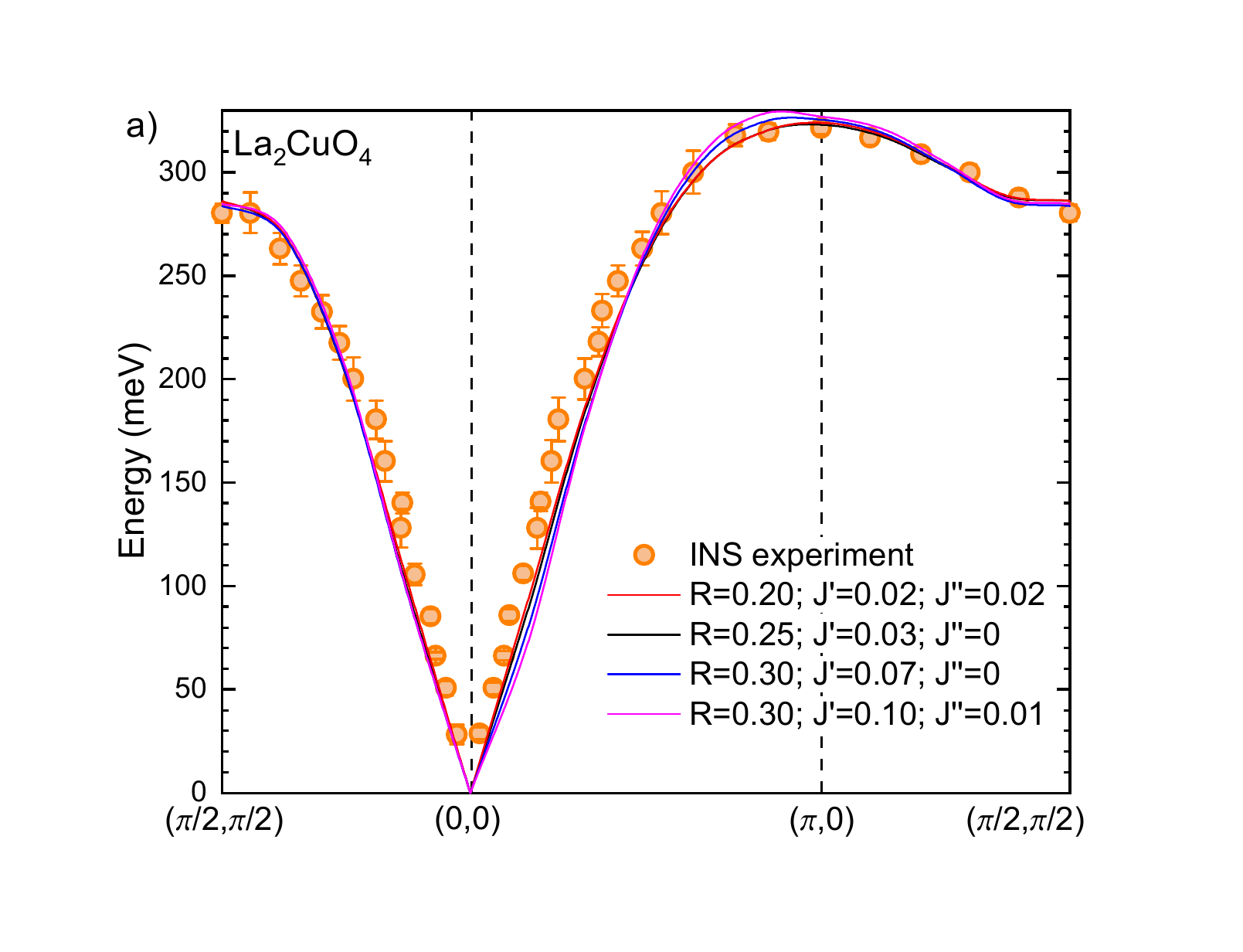}
\includegraphics[scale=0.4]{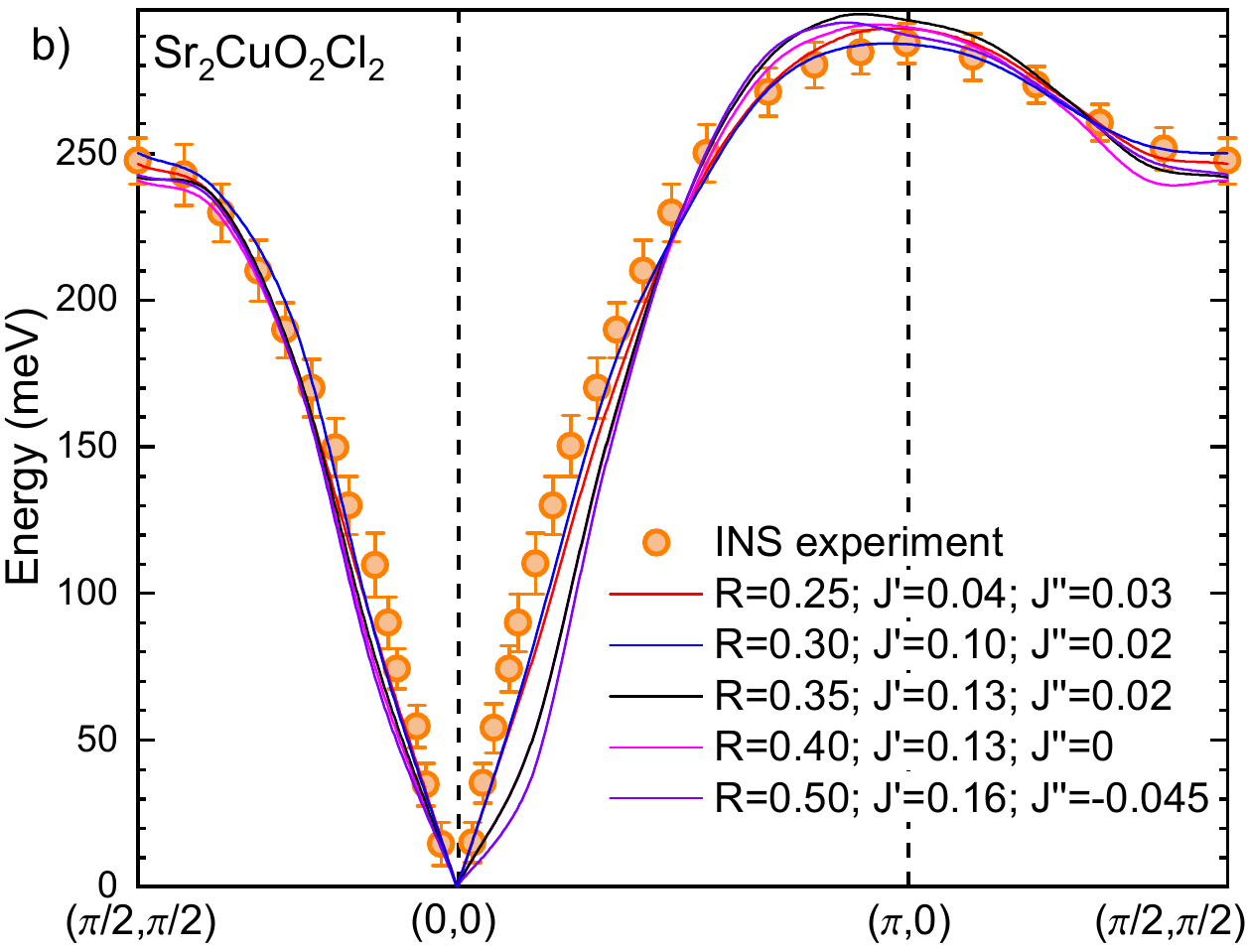}
\caption{Magnon dispersion in $\rm La_2CuO_4$ and $\rm Sr_2CuO_2Cl_2$ measured in inelastic neutron scattering (INS) experiments (Refs.~\cite{cupr3} and \cite{cupr1}, respectively) and calculated in the BOT using some sets of model parameters (see also Table~\ref{table1} and Fig.~\ref{bz}).
\label{magnon}}
\end{figure}

In agreement with many previous considerations, we find that there are many sets of model parameters which can fit reasonably good experimentally measured magnon spectra in layered cuprates. The result of our fitting is shown in Fig.~\ref{magnon} for $\rm La_2CuO_4$ and $\rm Sr_2CuO_2Cl_2$ and obtained values of $J$ for each set of parameters are summarized in Table~\ref{table1}. Parameters of the Hubbard model $U$ and $t$ estimated from Eqs.~\eqref{j} and \eqref{r} using $J$ and $R$ values are also listed in Table~\ref{table1}. 

\begin{table}
\caption{Some sets of parameters of model \eqref{ham} considered for application of our theory to $\rm La_2CuO_4$ and $\rm Sr_2CuO_2Cl_2$. Values of $J$ are obtained from the fitting of experimentally found magnon spectra (see Fig.~\ref{magnon}). Parameters of the Hubbard model $U$ and $t$ are estimated from Eqs.~\eqref{j} and \eqref{r} using $J$ and $R$ values. 
\label{table1}
}
\begin{tabular}{|c|c|c|c|c|c|c|}
\hline
& $R$ & $J'$ & $J''$ & $J$, meV & $U$, meV & $t$, meV \\
\hline
& $0.20J$  & $0.02J$ & $0.02J$ & 151.1 & $4244$ & $412$\\
$\rm La_2CuO_4$ & $0.25J$  & $0.03J$ & $0$ & 153.1 & $3539$ & $382$\\
& $0.30J$  & $0.07J$ & $0$ & 166.7 & $3302$ & $387$\\
& $0.30J$  & $0.10J$ & $0.01J$     & 175.0 & $3465$ & $406$\\
\hline
& $0.25J$  & $0.04J$ & $0.03J$ & 142.4 & $3292$ & $355$\\
& $0.30J$  & $0.10J$ & $0.02J$ & 158.9 & $3146$ & $369$\\
$\rm Sr_2CuO_2Cl_2$ & $0.35J$  & $0.13J$ & $0.02J$ & 172.8 & $3014$ & $379$\\
& $0.40J$  & $0.13J$ & $0$     & 171.5 & $2690$ & $359$\\
& $0.50J$  & $0.16J$ & $-0.045J$ & 175.5 & $2321$ & $342$\\
\hline
\end{tabular}
\end{table}

It is seen from Fig.~\ref{magnon}(a) that sets of parameters with $R=0.2J$ and $R=0.25J$ provide very good fits of experimental data in $\rm La_2CuO_4$ which are almost indistinguishable from each other. In both of these sets $J\approx150$~meV (see Table~\ref{table1}) that is in agreement with results of self-consistent spin-wave theory \cite{kataninmag}. The value of $R=0.24J$ proposed in Ref.~\cite{kataninmag} also does not contradict our findings. Notice that previous considerations based on linear spin-wave analysis underestimate $J$ value 10-15 \% and suggest considerably large $R\approx0.4J$ in $\rm La_2CuO_4$ (see, e.g., discussion in Refs.~\cite{katanin,kataninmag}). 

\begin{figure}
\includegraphics[scale=0.4]{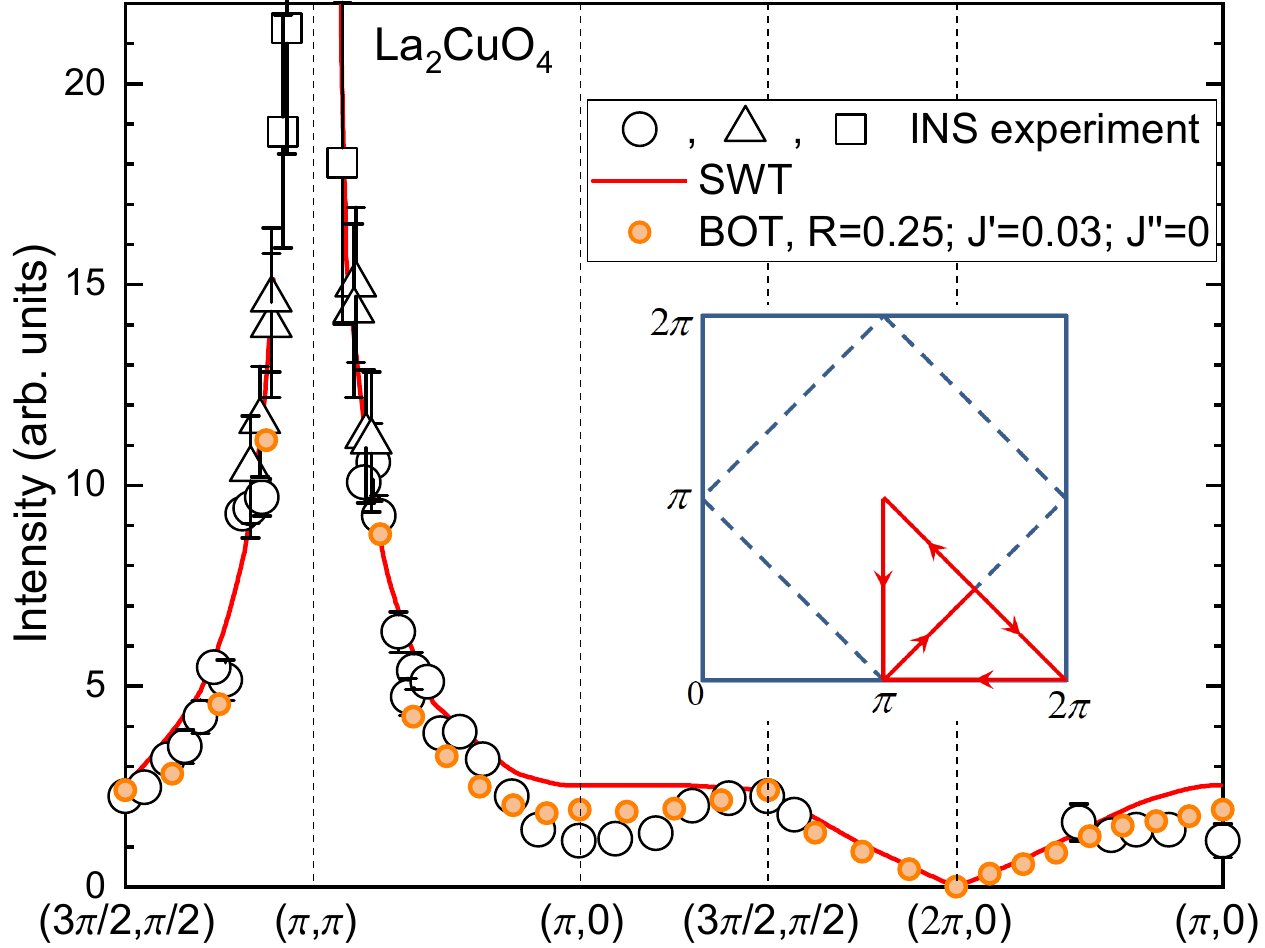}
\caption{Magnon spectral weights in $\rm La_2CuO_4$ measured in inelastic neutron scattering (INS) experiment \cite{cupr3}, found in Ref.~\cite{cupr3} within a spin-wave consideration (SWT), and calculated in the BOT using the set of model parameters with $R=0.25$ presented in Table~\ref{table1}. Each set of theoretical results (SWT and BOT) is multiplied by a common factor to fit experimental findings. INS and SWT data are taken from Fig.~2 of Ref.~\cite{cupr3}.
\label{weight}}
\end{figure}

Fig.~\ref{weight} shows that our calculations of magnon spectral weights using the set of parameters with $R=0.25J$ describe well corresponding experimental data in $\rm La_2CuO_4$. The small discrepancy near $(\pi,0)$ is assumed to be related with a higher-energy continuum of some excitations observed experimentally \cite{cupr3} which is very close to the magnon peak at $(\pi,0)$. The origin of this continuum is unknown now because it was not reproduced within model \eqref{ham} even in quantum Monte Carlo simulations (see Ref.~\cite{cupr3}).

Fig.~\ref{magnon}(b) shows that experimental data in $\rm Sr_2CuO_2Cl_2$ are reproduced well using sets of parameters with $R=0.25J$ and $R=0.3J$. Values of $t$ and $J$ found for the set with $R=0.25J$ and listed in Table~\ref{table1} are in agreement with values of $t\approx350$~meV and $J\approx140$~meV obtained in Ref.~\cite{arpes} by fitting ARPES data within the $t$--$J$ model.

As it is explained in the previous section, the amplitude mode does produce an anomaly in the longitudinal spin susceptibility $\chi_{zz}(\omega,{\bf k})$ \eqref{dsf}. However its small spectral weight hinders its detection in experiments with polarized neutrons (see Fig.~\ref{zzfig}). 

\subsection{Raman scattering in $B_{1g}$ geometry. Singlon.}

The operator of interaction of light with undoped cuprates in Raman scattering experiments can be derived in the large-$U$ Hubbard model as a series in parameter
\begin{equation}
\label{delta}
	\Delta = \frac{t}{U-\Omega},
\end{equation}
where $\Omega$ is the initial photon energy. \cite{ramansh} The result in the $B_{1g}$ geometry has the form in the second order in $\Delta$
\begin{equation}
\label{b1g}
H_{B_{1g}} \approx 
\left( \frac12 - 4\Delta^2 \right) 
\sum_{\bf r}{\bf S}_{\bf r}\left( {\bf S}_{\bf r+a} - {\bf S}_{\bf r+b} \right)
+
2\Delta^2 \sum_{\bf r}{\bf S}_{\bf r}\left( {\bf S}_{\bf r+{\rm2}a} - {\bf S}_{\bf r+{\rm2}b} \right)
+
8\Delta^2 \sum_{\bf r} \sum_{{\bf p},{\bf q}=\pm{\bf a},\pm{\bf b}}
i\epsilon_{\bf p,q} {\bf S}_{\bf r}\left[ {\bf S}_{\bf r+p} \times {\bf S}_{\bf r+q} \right],
\end{equation}
where $\bf a$ and $\bf b$ are square lattice translation vectors, $\epsilon_{\bf a,b}=1$ and $\epsilon_{\bf p,q}=-\epsilon_{\bf q,p}=-\epsilon_{-\bf p,q}$. \cite{ramansh} The intensity of signal in the Raman scattering experiment is proportional to $-{\rm Im}\chi_{PP}(\omega)$ given by Eq.~\eqref{dsf} with $P=H_{B_{1g}}$. The first term in Eq.~\eqref{b1g} is the famous Loudon-Fleury Hamiltonian \cite{lframan} which is proportional to $P^a_{\bf 0}-P^b_{\bf 0}$ (see Eqs.~\eqref{px} and \eqref{py}). Eq.~\eqref{b1g} is expected to work well when the photon energy is much smaller than the gap between the conducting and insulating bands. However it is not the case in most Raman experiments in cuprates because commonly used lasers with wavelengths 450--580 nm result in $\Delta\sim1$. A strong dependence of the Raman line shape on the energy of the incoming photon was observed experimentally in cuprates in the regime of $\Delta\sim1$. \cite{ramanexp4} Although the common Loudon-Fleury theory cannot account for this dependence, this theory is frequently used for the analysis of Raman data because it reproduces qualitatively many characteristic features of experimental results. At $\Delta\sim1$, a more accurate consideration of interaction of light with cuprates should be based on (extended) Hubbard models (see, e.g., Refs.~\cite{raman4,raman5,raman6}).

\begin{figure}
\includegraphics[scale=0.4]{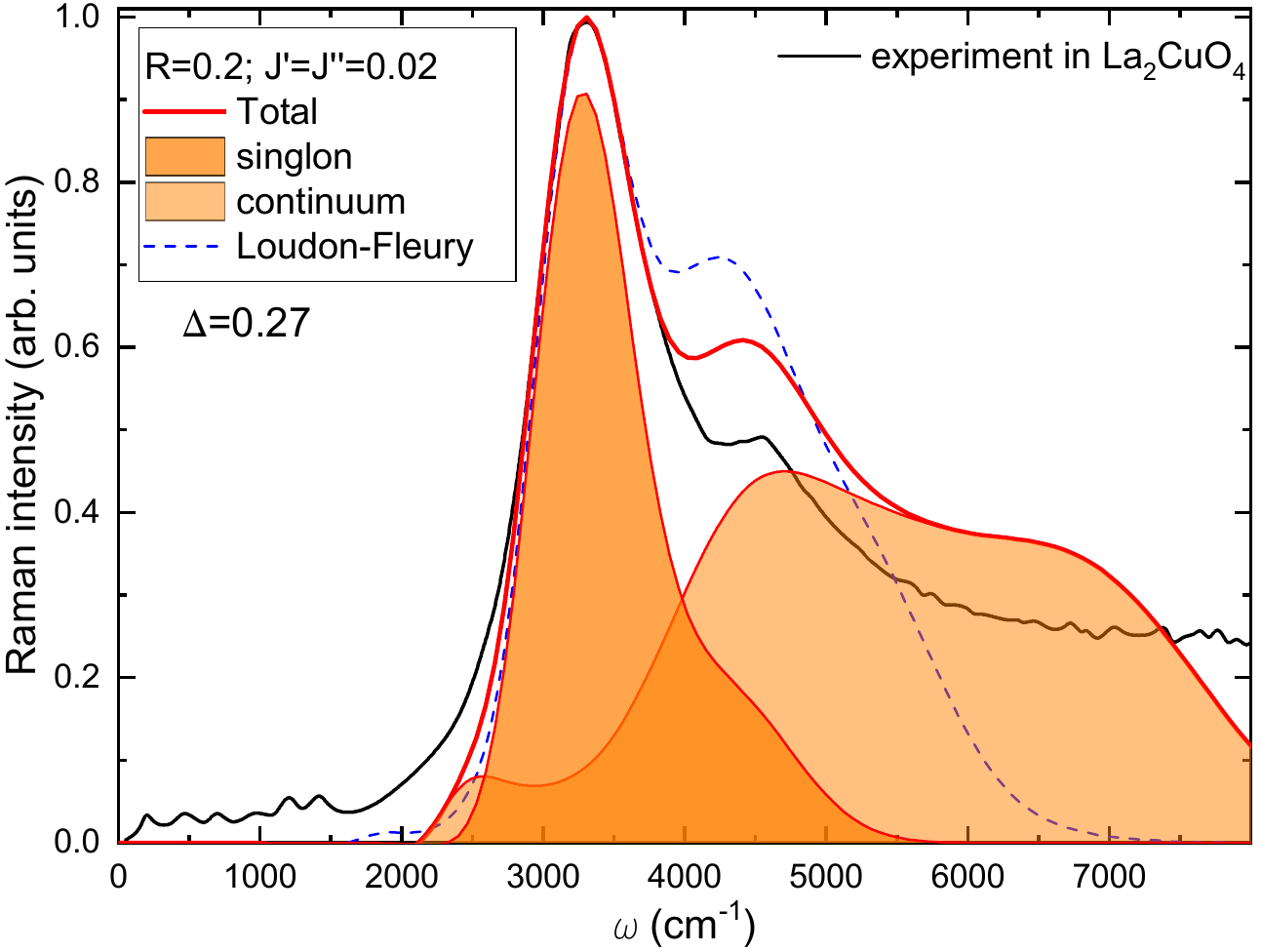}
\caption{Raman scattering intensity measured \cite{ramanexp3} in $\rm La_2CuO_4$ in the $B_{1g}$ geometry (experimental data are taken from Fig.~6(c) of Ref.~\cite{ramanexp3}). Theoretical results are also shown which are obtained in the BOT using Eq.~\eqref{b1g} and $\Delta=0.27$ as it is described in the text. The "two-magnon" peak at $\omega\approx3270$~cm$^{-1}$ is produced in our theory by spin-0 quasiparticle named singlon whereas the higher energy shoulder of the peak arises mainly from the diagram presented in Fig.~\ref{chifig}(d) and originates from two-particle continuum of spin-1 excitations. Both experimental and theoretical data are multiplied by common factors to normalize the peak height to unity. Theoretical results are also shown obtained in the BOT using the Loudon-Fleury Hamiltonian (the first term in Eq.~\eqref{b1g}).
\label{ramanfig}}
\end{figure}

As we obtained before \cite{ibot}, the Loudon-Fleury Hamiltonian in $B_{1g}$ geometry (the first term in Eq.~\eqref{b1g}) contains terms linear in Bose operators corresponding to singlon at zero momentum (linear terms do not arise in the BOT neither in other geometries, $A_{1g}$, $A_{2g}$, and $B_{2g}$, nor in the last two terms in Eq.~\eqref{b1g}). As a result, the generalized susceptibility describing Raman scattering contains Green's function of the singlon at $\bf k=0$ leading to a peak which was really observed experimentally and which is shown in Fig.~\ref{ramanfig}. In the spin-wave theory, this peak was obtained after the summation of ladder diagrams (that is why it is called "two-magnon peak" in the literature). \cite{raman1,raman2,raman3,raman4}

The only set of parameters listed in Table~\ref{table1} which might somehow satisfy the criterion of $\Delta\ll1$ is the set with the smallest $R=0.2J$ (which is relevant to $\rm La_2CuO_4$). For photon wavelength $\lambda=458$~nm used in experiments, $\Delta\approx0.27$ for this set of parameters whereas $\Delta>0.4$ for other sets. We calculate the Raman intensity within the BOT using Eq.~\eqref{b1g} with $\Delta=0.27$ and the set of parameters in Table~\ref{table1} with $R=0.2J$. As it was in previous numerical \cite{raman6,raman7} and analytical \cite{katanin,kataninmag} considerations, we obtain that the "two-magnon" peak at $\omega\approx3270$~cm$^{-1}$ is narrower than that in experimental data. The origin of the peak broadening has not been established. It may be related with the above mentioned fact that the scattering mechanism is not precisely known at $\Delta\sim1$. To fit experimental data, an artificial magnon damping is introduced "by hand" in analytical calculations (see, e.g., Ref.~\cite{katanin}) or a Gaussian average on $J$ values is performed (see, e.g., Ref.~\cite{raman6,raman7}). It is believed that this is a simple way to account for mechanisms not taken into account in purely spin models (in particular, the spin-lattice interaction, see Refs.~\cite{raman6,raman7} and references therein). Then, we also perform a Gaussian average of our theoretical data on $J$ values with the standard deviation $\sigma=0.1J$ to fit the peak width. We point out however that this procedure is not needed in description of the infrared light absorption considered below whose mechanism has been understood better. Then, the broad Raman signal is not seemingly related with the damping of the singlon whose contribution prevail to the narrow peak in the infrared light absorption (see below).

We plot in Fig.~\ref{ramanfig} our results together with experimental data for $\rm La_2CuO_4$ reported in Ref.~\cite{ramanexp3}. A good agreement is seen between the theory and experiment. In particular, the peak and its higher energy shoulder are reproduced well. We also show in Fig.~\ref{ramanfig} that results obtained within the BOT using only the first term in Eq.~\eqref{b1g} (i.e., in the Loudon-Fleury approximation) describe experimental data worse although they also reproduce the characteristic feature of experimental data: the peak and the shoulder. We find that the shoulder stems from the two-particle continuum of spin-1 excitations with a minor contribution of the continuum of two spin-0 quasiparticles. This should be contrasted with the conclusion of Ref.~\cite{weidinger2015} that the shoulder appears due to the scattering of photon on two amplitude excitations.

\subsection{RIXS with polarization analysis. Singlon and the amplitude mode.}

Resonant inelastic x-ray scattering (RIXS) has made remarkable progress as a spectroscopic technique in recent years. \cite{rixsrev} In cuprates and other transition metal compounds with large
superexchange coupling, it has allowed not only to reproduce magnon spectra known from the inelastic neutron scattering experiments but also to reveal anomalies attributed to excitations from $S_z=0$ sector (the so-called "bimagnon" and "continuum"). \cite{rixs1,rixs2,rixs3,rixs4,rixs5,rixs6,Tacon} The latter are manifested most clearly in experiments with analysis of the light polarization. \cite{rixs1,rixs6} Incoming photons are polarized either parallel ($\pi$) or perpendicular ($\sigma$) to the scattering plane. The rotation of the photon polarization as a result of the scattering implies a transfer of angular momentum and an excitation of quasiparticles in the sample with finite total spin projection. In contrast, the cross section of photons with no polarization changing (i.e., $\sigma\sigma'$ and $\pi\pi'$ channels, where the prime indicates the polarization of the scattered light) is governed by spin-0 quasiparticles and multi-quasiparticle processes having zero total spin projection. When $\sigma$ polarizations of incoming photons is directed along $y$ axis shown in Fig.~\ref{bz}, cross sections in $\sigma\sigma'$ and $\pi\pi'$ configurations are proportional to $-{\rm Im}\chi_{PP}(\omega)$ with $P=P^b_{\bf k}$ and $P^a_{\bf k}$ (see Eqs.~\eqref{dsfpp}--\eqref{py}), respectively (see, e.g., Ref.~\cite{rixs1}).

\begin{figure}
\includegraphics[scale=0.4]{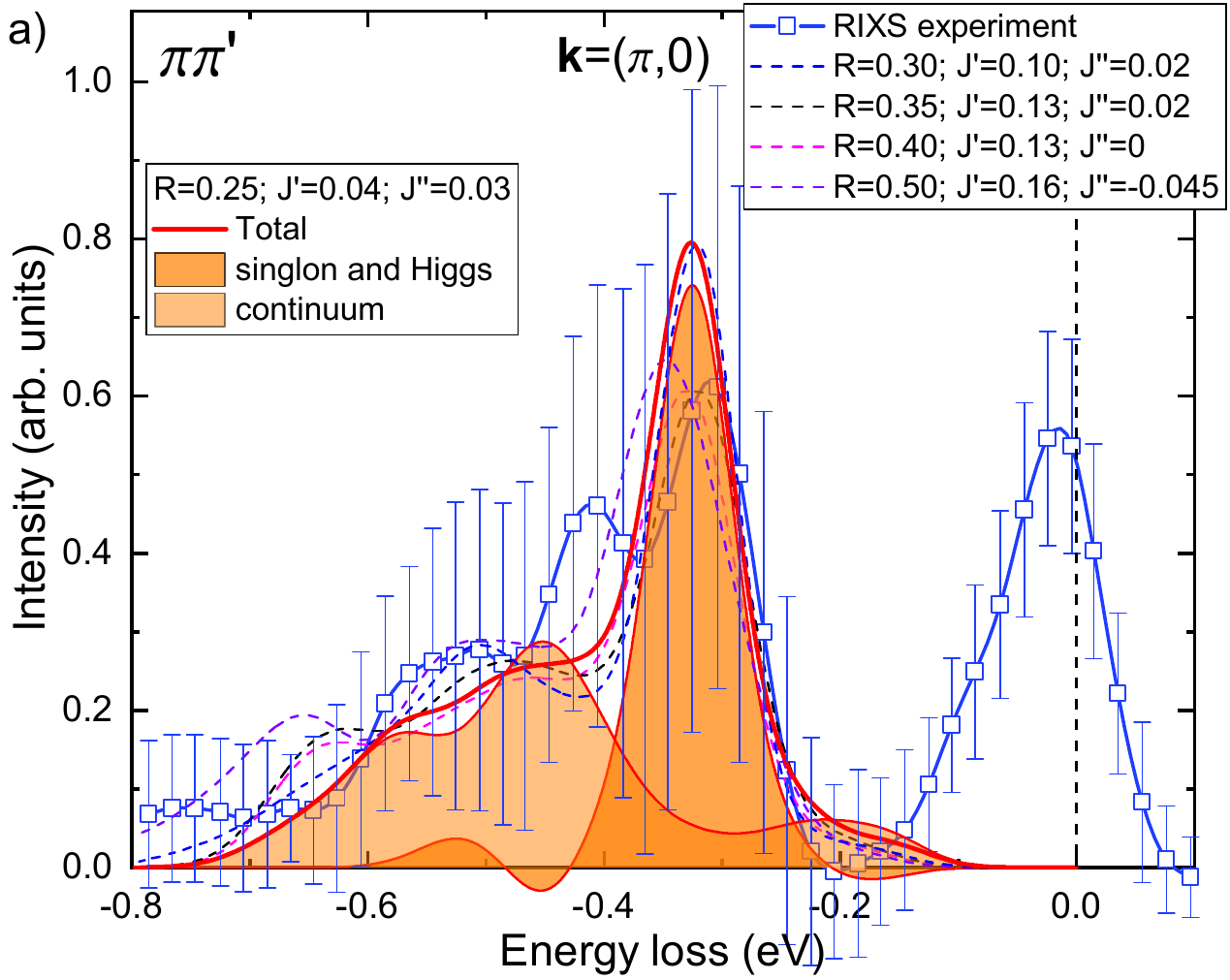}
\includegraphics[scale=0.4]{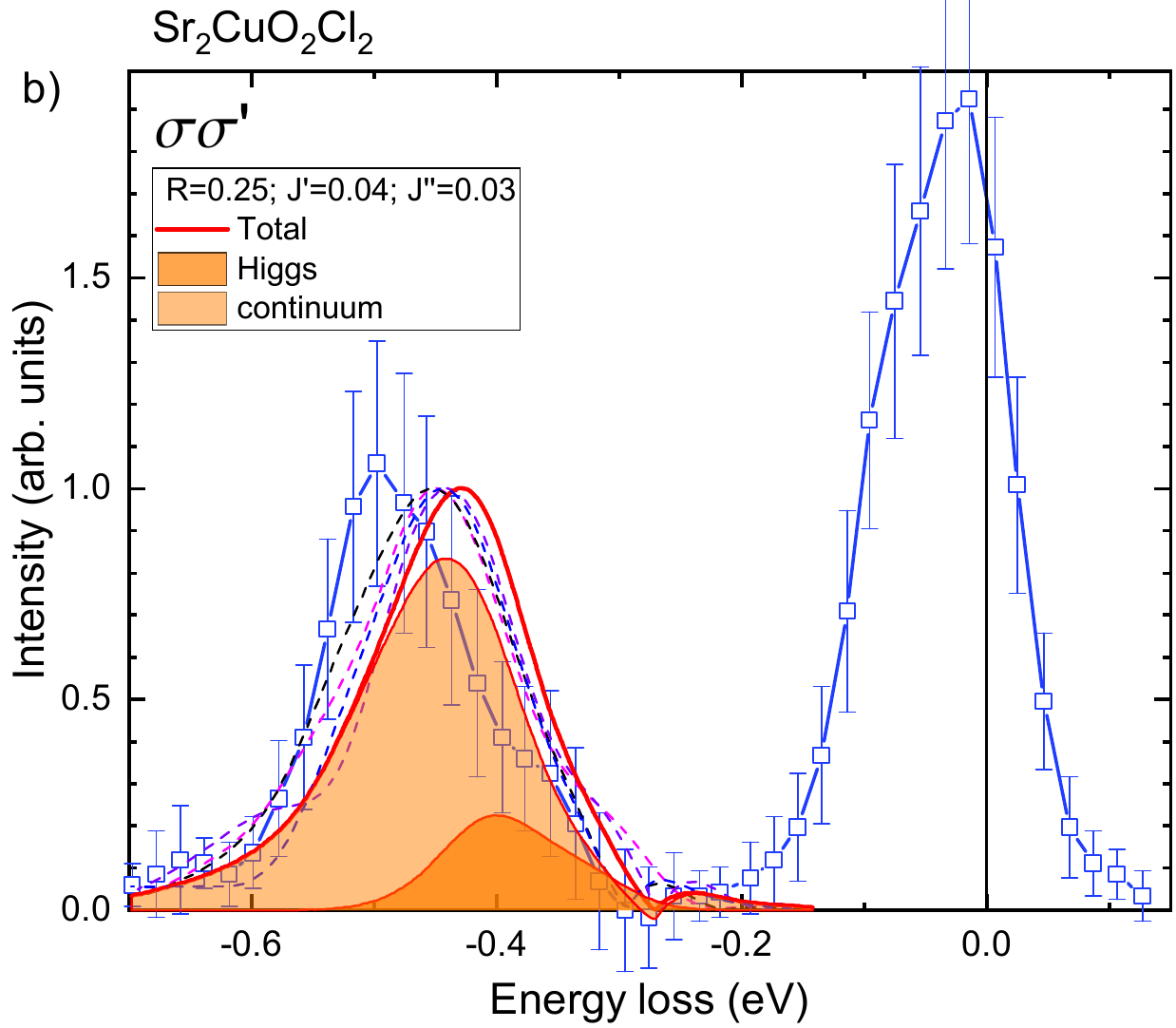}
\caption{Results of the RIXS experiment in $\rm Sr_2CuO_2Cl_2$ in (a) $\pi\pi'$ and (b) $\sigma\sigma'$ polarization at momentum ${\bf k}=(\pi,0)$ taken from Figs.~2(a), 2(b), and 3(b) of Ref.~\cite{rixs1}. Results are also shown of our calculations in the BOT using five sets of model parameters expressed in units of $J$ (see Table~\ref{table1}). Theoretical findings are convoluted with the experimental energy resolution of 32~meV. For the set with $R=0.25$, contributions to peaks from singlon and/or the amplitude (Higgs) excitations and from the two-particle continuum are distinguished by color.
\label{ppfig}}
\end{figure}

We show in Fig.~\ref{ppfig} experimental findings obtained \cite{rixs1} in $\sigma\sigma'$ and $\pi\pi'$ configurations in $\rm Sr_2CuO_2Cl_2$ at ${\bf k}=(\pi,0)$. Our results for $-{\rm Im}\chi_{PP}(\omega)$ with $P=P^b_{\bf k}$ and $P^a_{\bf k}$ are also presented in Fig.~\ref{ppfig} which are actually data shown in Figs.~\ref{ppfigs}(l) and \ref{ppfigs}(d), respectively, and convoluted with the experimental energy resolution. Because the available experimental data are quite noisy, a good agreement is seen between the experiment and our theory for almost all sets of model parameters listed in Table~\ref{table1}. It is shown in Fig.~\ref{ppfig}(a) that in the $\pi\pi'$ configuration the singlon and the Higss mode are responsible for the first experimental peak whereas the higher energy structure is produced by the continuum originating from diagrams depicted in Fig.~\ref{chifig}(b) and \ref{chifig}(d). In the $\sigma\sigma'$ channel presented in Fig.~\ref{ppfig}(b), the broad peak is a sum of the continuum and the anomaly produced by the amplitude mode. It is seen from Figs.~\ref{ppfigs}(l) and \ref{ppfigs}(d) that better experimental energy resolution and better accuracy in intensity measurement would allow to distinguish contributions of the singlon and the Higgs mode from anomalies originating from edges of two-particle continuums. The major contribution to continuums in both configurations is made by two spin-1 quasiparticles. 

\subsection{Infrared spectroscopy. Singlon and the amplitude mode.}

\begin{figure}
\includegraphics[scale=0.4]{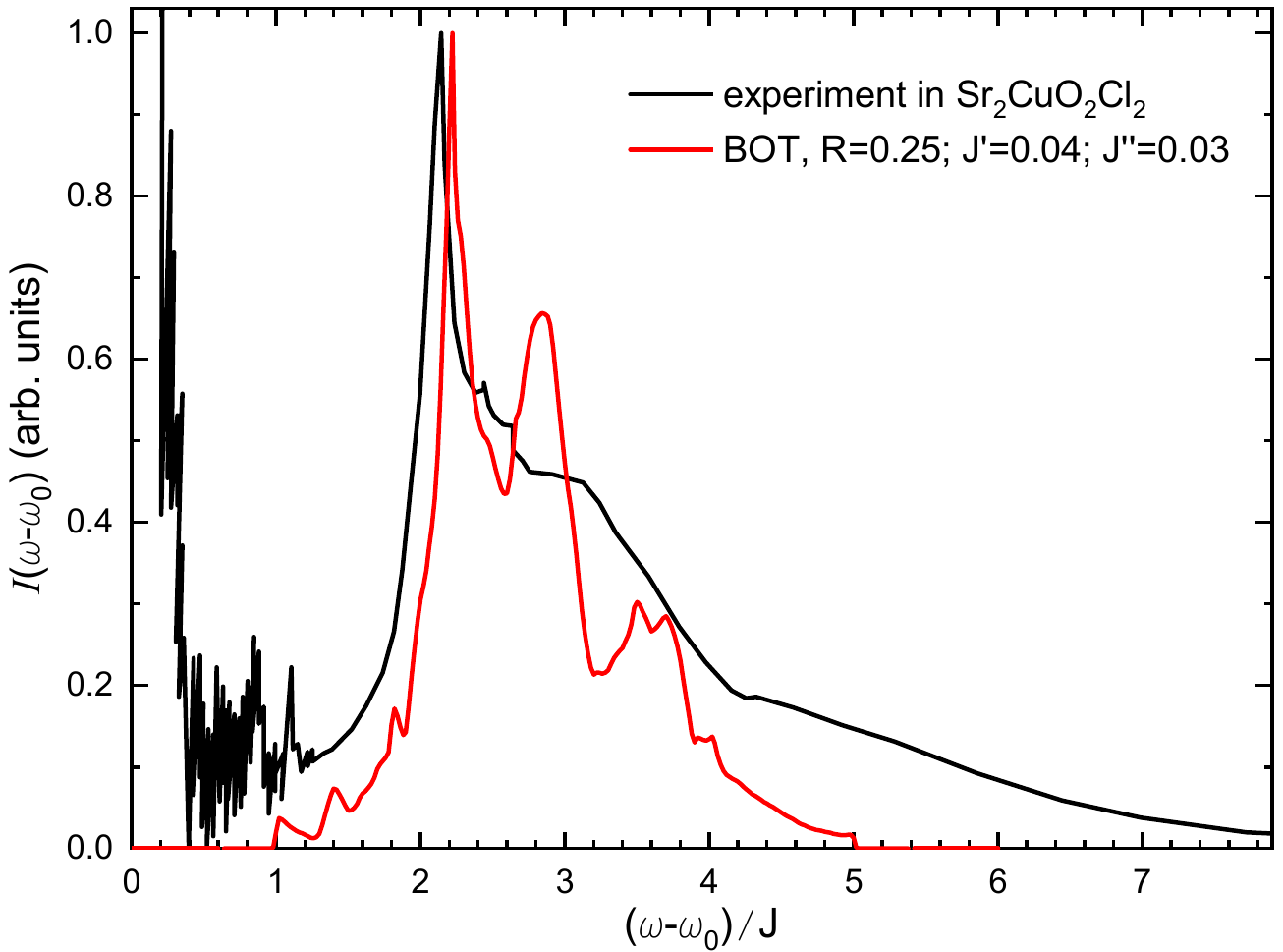}
\caption{$I(\omega-\omega_0)$ extracted using Eq.~\eqref{alpha} from infrared optical absorption spectra in $\rm Sr_2CuO_2Cl_2$ and taken from Fig.~1 of Ref.~\cite{ir2}. Also shown are results of our calculation within the BOT of $I(\omega-\omega_0)$ given by Eq.~\eqref{i}. Both sets of data are multiplied by common factors to normalize the peak height to unity.
\label{irfig}}
\end{figure}

The theory of infrared optical absorption in insulating magnetic materials is developed in Refs.~\cite{ir1,ir2,ir3,ir4,ir5}. The mechanism of light absorption is well understood and includes a simultaneous creation of a phonon and a multimagnon excitation. The absorption spectrum $\alpha(\omega)$ has the form \cite{ir2,ir3,ir4,ir5}
\begin{eqnarray}
\label{alpha}
\alpha(\omega) &=& \alpha_0\omega I(\omega-\omega_0),\\
\label{i}
I(\omega) &=& -\frac{8}{\pi N}
\sum_{\bf k}\sin^2\frac{k_a}{2} \left( \sin^2\frac{k_a}{2} + \sin^2\frac{k_b}{2} \right)
{\rm Im} \chi_{PP}(\omega),
\end{eqnarray}
where $\alpha_0$ is a material dependent constant, $\omega_0$ is the frequency of the stretching mode
phonon, $\chi_{PP}(\omega)$ is given by Eq.~\eqref{dsfpp}, $P=P^a_{\bf k}$ (see Eq.~\eqref{px}), and the summation is performed over the crystal BZ (see Fig.~\ref{bz}). Although $I(\omega)$ given by Eq.~\eqref{i} is an integral characteristic, the experimentally obtained absorption spectrum contains a sharp anomaly shown in Fig.~\ref{irfig} which is attributed in Refs.~\cite{ir2,ir3} to two-magnon bound states ("bimagnons") with momenta close to $(\pi,0)$.

The result of our calculation of $I(\omega)$ within the BOT is also presented in Fig.~\ref{irfig} which reproduce well the sharp anomaly near $2.2J$. 
\footnote{
We calculate ${\rm Im} \chi_{PP}(\omega)$ on a grid $10\times10$ in the BZ to find the sum in Eq.~\eqref{i}.
}
We confirm the conclusion of Refs.~\cite{ir2,ir3} that this peak is produced by well defined two-magnon bound states with momenta close to $(\pi,0)$ (this can be seen from Fig.~\ref{ppfigs} taking into account that the intensity scales are the largest in Figs.~\ref{ppfigs}(c)--(e) and that momenta with $k_a\approx\pi$ have the largest weights in the sum in Eq.~\eqref{i} due to the factor $\sin^2(k_a/2)$). The only clarification we make is that {\it two} well-defined spin-0 quasiparticles with momenta close to $(\pi,0)$ contribute to this anomaly, the singlon and the amplitude mode, and that the contribution to the peak (the spectral weight) of the singlon is about five times larger than that of the Higgs mode.  Besides, our estimations above of $J$ value in $\rm Sr_2CuO_2Cl_2$ is 15--20\% larger than the value proposed in Refs.~\cite{ir2,ir3} from the analisys of the infrared absorption in the spin-wave theory. Notice also that in contrast to the consideration of the Raman scattering carried out above, we do not perform any Gaussian averaging of our results obtained in the BOT in order to fit experimental data.

It is seen from Fig.~\ref{irfig} that the BOT predicts a broad peak near $2.9J$. As it can be inferred from Fig.~\ref{ppfigs}, this peak originates from Higgs excitations with momenta close to $(\pi,\pi)$ and from the continuum of two spin-1 quasiparticles. However experimental data show a shoulder-like anomaly in that region instead of the broad peak with approximately the same spectral weight. This discrepancy may originate from an underestimation of the amplitude mode damping in the first order in $1/n$. The structure around $3.7J$ in Fig.~\ref{irfig} obtained in the BOT stems mainly from the continuum of two spin-1 excitations. Other very small peaks on the theoretical curve in Fig.~\ref{irfig} do not correspond to quasiparticles and stem from the incoherent background which has rather sharp low anomalies in some points of the BZ seen in Fig.~\ref{ppfigs}.

\section{Summary and conclusion}
\label{conc}

To conclude, we discuss using the bond-operator theory (BOT) low-energy elementary excitations in the Heisenberg spin-$\frac12$ antiferromagnet \eqref{ham} with the ring exchange and small interactions between the second- and the third-neighbor spins on the square lattice at $T=0$. This model was suggested before for description of parent compounds of high-temperature superconducting layered cuprates. We demonstrate that in addition to magnons (spin-1 excitations) there are three well-defined spin-0 quasiparticles whose energies lie near the magnon spectrum (see Fig.~\ref{spectra}). Two of them, the amplitude (Higgs) mode and the quasiparticle which we call singlon, produce pronounced anomalies in experimental data on the Raman scattering, resonant inelastic x-ray scattering (RIXS), and infrared optical absorption. BOT proved to be a powerful, precise, and convenient tool for discussion of some salient features of spin dynamics whose interpretation is labored in conventional theoretical approaches. Both spin-1 and spin-0 excitations are described by separate bosons in the BOT that allows to find their spectra and their contributions to observable quantities on equal footing by considering the same diagrams.

In agreement with previous results, we demonstrate that a phase transition arises at $R\agt J$ in model \eqref{ham} at which the magnon spectrum becomes zero at momentum ${\bf k}=(\pi/2,\pi/2)$. Presumably, this transition is accompanied also with a "condensation" of the spin-0 mode shown in Fig.~\ref{spectra} in green. Our results on the magnon spectrum are in agreement with previous findings obtained using the exact diagonalization of finite clusters (see Fig.~\ref{mag2}).

We fit the magnon dispersion and magnon spectral weights obtained experimentally in $\rm La_2CuO_4$ and $\rm Sr_2CuO_2Cl_2$ and point out optimal model parameters (see Figs.~\ref{magnon}, \ref{weight}, and first two lines for each substance in Table~\ref{table1}). They are in a good agreement with previous results obtained using the self-consistent spin-wave theory (in $\rm La_2CuO_4$) and $t$--$J$ model (in $\rm Sr_2CuO_2Cl_2$). We find that although the amplitude mode produces an anomaly in the longitudinal spin susceptibility $\chi_{zz}(\omega,{\bf k})$ given by Eq.~\eqref{dsf}, its small spectral weight hinders its determination in polarized neutron scattering experiments (see Fig.~\ref{zzfig}).

We show that the singlon with zero momentum produces a distinct peak in the Raman scattering in the $B_{1g}$ symmetry that is known in the literature as a "two-magnon" peak. The higher energy shoulder in the Raman signal arises due to the incoherent two-particle continuum formed predominantly by spin-1 excitations (see Fig.~\ref{ramanfig} for $\rm La_2CuO_4$).

We demonstrate that the singlon and the amplitude mode can be probed in RIXS experiments with polarization analysis in $\pi\pi'$ and $\sigma\sigma'$ configurations. These quasiparticles produce noticeable anomalies in the cross sections which are clearly seen in many parts of the BZ against the structured two-particle continuum (see Fig.~\ref{ppfigs}). Our particular calculations for $\rm Sr_2CuO_2Cl_2$ at ${\bf k}=(\pi,0)$ are in agreement with recent experimental data shown in Fig.~\ref{ppfig}.

We demonstrate that singlons and amplitude excitations with momenta close to $(\pi,0)$ produce the sharp peak in the infrared optical absorption spectra in $\rm Sr_2CuO_2Cl_2$ presented in Fig.~\ref{irfig} whereas the higher energy shoulder appears due to the continuum and the Higgs excitations with momenta lying near $(\pi,\pi)$.

It is a natural idea to generalize the suggested approach to strongly correlated electron systems with charge carriers. In particular, it looks promising because of the great role of short-range spin correlations and the small size of Cooper pairs in high-temperature superconductors. Such approach will be presented in forthcoming papers.


\begin{thebibliography}{56}
\expandafter\ifx\csname natexlab\endcsname\relax\def\natexlab#1{#1}\fi
\expandafter\ifx\csname bibnamefont\endcsname\relax
  \def\bibnamefont#1{#1}\fi
\expandafter\ifx\csname bibfnamefont\endcsname\relax
  \def\bibfnamefont#1{#1}\fi
\expandafter\ifx\csname citenamefont\endcsname\relax
  \def\citenamefont#1{#1}\fi
\expandafter\ifx\csname url\endcsname\relax
  \def\url#1{\texttt{#1}}\fi
\expandafter\ifx\csname urlprefix\endcsname\relax\def\urlprefix{URL }\fi
\providecommand{\bibinfo}[2]{#2}
\providecommand{\eprint}[2][]{\url{#2}}

\bibitem[{\citenamefont{{Manousakis}}(1991)}]{monous}
\bibinfo{author}{\bibfnamefont{E.}~\bibnamefont{{Manousakis}}},
  \bibinfo{journal}{Reviews of Modern Physics} \textbf{\bibinfo{volume}{63}},
  \bibinfo{pages}{1} (\bibinfo{year}{1991}).

\bibitem[{\citenamefont{Keimer et~al.}(2015)\citenamefont{Keimer, Kivelson,
  Norman, Uchida, and Zaanen}}]{Keimer2015}
\bibinfo{author}{\bibfnamefont{B.}~\bibnamefont{Keimer}},
  \bibinfo{author}{\bibfnamefont{S.~A.} \bibnamefont{Kivelson}},
  \bibinfo{author}{\bibfnamefont{M.~R.} \bibnamefont{Norman}},
  \bibinfo{author}{\bibfnamefont{S.}~\bibnamefont{Uchida}}, \bibnamefont{and}
  \bibinfo{author}{\bibfnamefont{J.}~\bibnamefont{Zaanen}},
  \bibinfo{journal}{Nature} \textbf{\bibinfo{volume}{518}},
  \bibinfo{pages}{179} (\bibinfo{year}{2015}).

\bibitem[{\citenamefont{{Anderson}}(1987)}]{anderson}
\bibinfo{author}{\bibfnamefont{P.~W.} \bibnamefont{{Anderson}}},
  \bibinfo{journal}{Science} \textbf{\bibinfo{volume}{235}},
  \bibinfo{pages}{1196} (\bibinfo{year}{1987}).

\bibitem[{\citenamefont{Auerbach}(1994)}]{auer}
\bibinfo{author}{\bibfnamefont{A.}~\bibnamefont{Auerbach}},
  \emph{\bibinfo{title}{Interacting Electrons and Quantum Magnetism}}
  (\bibinfo{publisher}{Springer, New York}, \bibinfo{year}{1994}).

\bibitem[{\citenamefont{Reischl et~al.}(2004)\citenamefont{Reischl,
  M\"uller-Hartmann, and Uhrig}}]{hubb1}
\bibinfo{author}{\bibfnamefont{A.}~\bibnamefont{Reischl}},
  \bibinfo{author}{\bibfnamefont{E.}~\bibnamefont{M\"uller-Hartmann}},
  \bibnamefont{and} \bibinfo{author}{\bibfnamefont{G.~S.} \bibnamefont{Uhrig}},
  \bibinfo{journal}{Phys. Rev. B} \textbf{\bibinfo{volume}{70}},
  \bibinfo{pages}{245124} (\bibinfo{year}{2004}).

\bibitem[{\citenamefont{Coldea et~al.}(2001)\citenamefont{Coldea, Hayden,
  Aeppli, Perring, Frost, Mason, Cheong, and Fisk}}]{cupr5}
\bibinfo{author}{\bibfnamefont{R.}~\bibnamefont{Coldea}},
  \bibinfo{author}{\bibfnamefont{S.~M.} \bibnamefont{Hayden}},
  \bibinfo{author}{\bibfnamefont{G.}~\bibnamefont{Aeppli}},
  \bibinfo{author}{\bibfnamefont{T.~G.} \bibnamefont{Perring}},
  \bibinfo{author}{\bibfnamefont{C.~D.} \bibnamefont{Frost}},
  \bibinfo{author}{\bibfnamefont{T.~E.} \bibnamefont{Mason}},
  \bibinfo{author}{\bibfnamefont{S.-W.} \bibnamefont{Cheong}},
  \bibnamefont{and} \bibinfo{author}{\bibfnamefont{Z.}~\bibnamefont{Fisk}},
  \bibinfo{journal}{Phys. Rev. Lett.} \textbf{\bibinfo{volume}{86}},
  \bibinfo{pages}{5377} (\bibinfo{year}{2001}).

\bibitem[{\citenamefont{Delannoy et~al.}(2009)\citenamefont{Delannoy, Gingras,
  Holdsworth, and Tremblay}}]{Uttt}
\bibinfo{author}{\bibfnamefont{J.-Y.~P.} \bibnamefont{Delannoy}},
  \bibinfo{author}{\bibfnamefont{M.~J.~P.} \bibnamefont{Gingras}},
  \bibinfo{author}{\bibfnamefont{P.~C.~W.} \bibnamefont{Holdsworth}},
  \bibnamefont{and} \bibinfo{author}{\bibfnamefont{A.-M.~S.}
  \bibnamefont{Tremblay}}, \bibinfo{journal}{Phys. Rev. B}
  \textbf{\bibinfo{volume}{79}}, \bibinfo{pages}{235130}
  (\bibinfo{year}{2009}).

\bibitem[{\citenamefont{Roger and Delrieu}(1989)}]{ring1}
\bibinfo{author}{\bibfnamefont{M.}~\bibnamefont{Roger}} \bibnamefont{and}
  \bibinfo{author}{\bibfnamefont{J.~M.} \bibnamefont{Delrieu}},
  \bibinfo{journal}{Phys. Rev. B} \textbf{\bibinfo{volume}{39}},
  \bibinfo{pages}{2299} (\bibinfo{year}{1989}).

\bibitem[{\citenamefont{Schmidt and Kuramoto}(1990)}]{ring2}
\bibinfo{author}{\bibfnamefont{H.}~\bibnamefont{Schmidt}} \bibnamefont{and}
  \bibinfo{author}{\bibfnamefont{Y.}~\bibnamefont{Kuramoto}},
  \bibinfo{journal}{Physica C: Superconductivity}
  \textbf{\bibinfo{volume}{167}}, \bibinfo{pages}{263} (\bibinfo{year}{1990}).

\bibitem[{\citenamefont{Headings et~al.}(2010)\citenamefont{Headings, Hayden,
  Coldea, and Perring}}]{cupr3}
\bibinfo{author}{\bibfnamefont{N.~S.} \bibnamefont{Headings}},
  \bibinfo{author}{\bibfnamefont{S.~M.} \bibnamefont{Hayden}},
  \bibinfo{author}{\bibfnamefont{R.}~\bibnamefont{Coldea}}, \bibnamefont{and}
  \bibinfo{author}{\bibfnamefont{T.~G.} \bibnamefont{Perring}},
  \bibinfo{journal}{Phys. Rev. Lett.} \textbf{\bibinfo{volume}{105}},
  \bibinfo{pages}{247001} (\bibinfo{year}{2010}).

\bibitem[{\citenamefont{Plumb et~al.}(2014)\citenamefont{Plumb, Savici,
  Granroth, Chou, and Kim}}]{cupr1}
\bibinfo{author}{\bibfnamefont{K.~W.} \bibnamefont{Plumb}},
  \bibinfo{author}{\bibfnamefont{A.~T.} \bibnamefont{Savici}},
  \bibinfo{author}{\bibfnamefont{G.~E.} \bibnamefont{Granroth}},
  \bibinfo{author}{\bibfnamefont{F.~C.} \bibnamefont{Chou}}, \bibnamefont{and}
  \bibinfo{author}{\bibfnamefont{Y.-J.} \bibnamefont{Kim}},
  \bibinfo{journal}{Phys. Rev. B} \textbf{\bibinfo{volume}{89}},
  \bibinfo{pages}{180410} (\bibinfo{year}{2014}).

\bibitem[{\citenamefont{Katanin and Kampf}(2002)}]{kataninmag}
\bibinfo{author}{\bibfnamefont{A.~A.} \bibnamefont{Katanin}} \bibnamefont{and}
  \bibinfo{author}{\bibfnamefont{A.~P.} \bibnamefont{Kampf}},
  \bibinfo{journal}{Phys. Rev. B} \textbf{\bibinfo{volume}{66}},
  \bibinfo{pages}{100403} (\bibinfo{year}{2002}).

\bibitem[{\citenamefont{Lorenzana and Sawatzky}(1995{\natexlab{a}})}]{ir2}
\bibinfo{author}{\bibfnamefont{J.}~\bibnamefont{Lorenzana}} \bibnamefont{and}
  \bibinfo{author}{\bibfnamefont{G.~A.} \bibnamefont{Sawatzky}},
  \bibinfo{journal}{Phys. Rev. Lett.} \textbf{\bibinfo{volume}{74}},
  \bibinfo{pages}{1867} (\bibinfo{year}{1995}{\natexlab{a}}).

\bibitem[{\citenamefont{Lorenzana and Sawatzky}(1995{\natexlab{b}})}]{ir3}
\bibinfo{author}{\bibfnamefont{J.}~\bibnamefont{Lorenzana}} \bibnamefont{and}
  \bibinfo{author}{\bibfnamefont{G.~A.} \bibnamefont{Sawatzky}},
  \bibinfo{journal}{Phys. Rev. B} \textbf{\bibinfo{volume}{52}},
  \bibinfo{pages}{9576} (\bibinfo{year}{1995}{\natexlab{b}}).

\bibitem[{\citenamefont{Katanin and Kampf}(2003)}]{katanin}
\bibinfo{author}{\bibfnamefont{A.~A.} \bibnamefont{Katanin}} \bibnamefont{and}
  \bibinfo{author}{\bibfnamefont{A.~P.} \bibnamefont{Kampf}},
  \bibinfo{journal}{Phys. Rev. B} \textbf{\bibinfo{volume}{67}},
  \bibinfo{pages}{100404} (\bibinfo{year}{2003}).

\bibitem[{\citenamefont{Sim\'on et~al.}(1996)\citenamefont{Sim\'on, Aligia,
  Batista, Gagliano, and Lema}}]{raman6}
\bibinfo{author}{\bibfnamefont{M.~E.} \bibnamefont{Sim\'on}},
  \bibinfo{author}{\bibfnamefont{A.~A.} \bibnamefont{Aligia}},
  \bibinfo{author}{\bibfnamefont{C.~D.} \bibnamefont{Batista}},
  \bibinfo{author}{\bibfnamefont{E.~R.} \bibnamefont{Gagliano}},
  \bibnamefont{and} \bibinfo{author}{\bibfnamefont{F.}~\bibnamefont{Lema}},
  \bibinfo{journal}{Phys. Rev. B} \textbf{\bibinfo{volume}{54}},
  \bibinfo{pages}{R3780} (\bibinfo{year}{1996}).

\bibitem[{\citenamefont{Majumdar et~al.}(2012)\citenamefont{Majumdar, Furton,
  and Uhrig}}]{ringth}
\bibinfo{author}{\bibfnamefont{K.}~\bibnamefont{Majumdar}},
  \bibinfo{author}{\bibfnamefont{D.}~\bibnamefont{Furton}}, \bibnamefont{and}
  \bibinfo{author}{\bibfnamefont{G.~S.} \bibnamefont{Uhrig}},
  \bibinfo{journal}{Phys. Rev. B} \textbf{\bibinfo{volume}{85}},
  \bibinfo{pages}{144420} (\bibinfo{year}{2012}).

\bibitem[{\citenamefont{Syromyatnikov}(2018)}]{ibot}
\bibinfo{author}{\bibfnamefont{A.~V.} \bibnamefont{Syromyatnikov}},
  \bibinfo{journal}{Phys. Rev. B} \textbf{\bibinfo{volume}{98}},
  \bibinfo{pages}{184421} (\bibinfo{year}{2018}).

\bibitem[{\citenamefont{Syromyatnikov and Aktersky}(2019)}]{aktersky}
\bibinfo{author}{\bibfnamefont{A.~V.} \bibnamefont{Syromyatnikov}}
  \bibnamefont{and} \bibinfo{author}{\bibfnamefont{A.~Y.}
  \bibnamefont{Aktersky}}, \bibinfo{journal}{Phys. Rev. B}
  \textbf{\bibinfo{volume}{99}}, \bibinfo{pages}{224402}
  (\bibinfo{year}{2019}).

\bibitem[{\citenamefont{Syromyatnikov}(2020)}]{iboth}
\bibinfo{author}{\bibfnamefont{A.~V.} \bibnamefont{Syromyatnikov}},
  \bibinfo{journal}{Phys. Rev. B} \textbf{\bibinfo{volume}{102}},
  \bibinfo{pages}{014409} (\bibinfo{year}{2020}).

\bibitem[{\citenamefont{Syromyatnikov}(2022{\natexlab{a}})}]{itri}
\bibinfo{author}{\bibfnamefont{A.~V.} \bibnamefont{Syromyatnikov}},
  \bibinfo{journal}{Phys. Rev. B} \textbf{\bibinfo{volume}{105}},
  \bibinfo{pages}{144414} (\bibinfo{year}{2022}{\natexlab{a}}).

\bibitem[{\citenamefont{Syromyatnikov}(2023)}]{itrih}
\bibinfo{author}{\bibfnamefont{A.}~\bibnamefont{Syromyatnikov}},
  \bibinfo{journal}{Annals of Physics} \textbf{\bibinfo{volume}{454}},
  \bibinfo{pages}{169342} (\bibinfo{year}{2023}).

\bibitem[{\citenamefont{Syromyatnikov}(2022{\natexlab{b}})}]{itrij1j2}
\bibinfo{author}{\bibfnamefont{A.~V.} \bibnamefont{Syromyatnikov}},
  \bibinfo{journal}{Phys. Rev. B} \textbf{\bibinfo{volume}{106}},
  \bibinfo{pages}{184415} (\bibinfo{year}{2022}{\natexlab{b}}).

\bibitem[{sup()}]{supp2}
\bibinfo{note}{See Supplemental Material at
  http://link.aps.org/supplemental/10.1103/PhysRevB.106.184415 for the code in
  the Mathematica software which generates the spin representation which we use
  in the present work for consideration of the stripe phase.}

\bibitem[{\citenamefont{Igarashi}(1992)}]{igar}
\bibinfo{author}{\bibfnamefont{J.-i.} \bibnamefont{Igarashi}},
  \bibinfo{journal}{Phys. Rev. B} \textbf{\bibinfo{volume}{46}},
  \bibinfo{pages}{10763} (\bibinfo{year}{1992}).

\bibitem[{\citenamefont{{Igarashi} and {Nagao}}(2005)}]{igar2}
\bibinfo{author}{\bibfnamefont{J.-I.} \bibnamefont{{Igarashi}}}
  \bibnamefont{and} \bibinfo{author}{\bibfnamefont{T.}~\bibnamefont{{Nagao}}},
  \bibinfo{journal}{\prb} \textbf{\bibinfo{volume}{72}},
  \bibinfo{pages}{014403} (\bibinfo{year}{2005}).

\bibitem[{\citenamefont{{Syromyatnikov}}(2010)}]{syromyat}
\bibinfo{author}{\bibfnamefont{A.~V.} \bibnamefont{{Syromyatnikov}}},
  \bibinfo{journal}{Journal of Physics: Condensed Matter}
  \textbf{\bibinfo{volume}{22}}, \bibinfo{pages}{216003}
  (\bibinfo{year}{2010}).

\bibitem[{\citenamefont{{Zheng} et~al.}(2005)\citenamefont{{Zheng}, {Oitmaa},
  and {Hamer}}}]{series}
\bibinfo{author}{\bibfnamefont{W.}~\bibnamefont{{Zheng}}},
  \bibinfo{author}{\bibfnamefont{J.}~\bibnamefont{{Oitmaa}}}, \bibnamefont{and}
  \bibinfo{author}{\bibfnamefont{C.~J.} \bibnamefont{{Hamer}}},
  \bibinfo{journal}{Phys. Rev. B} \textbf{\bibinfo{volume}{71}},
  \bibinfo{pages}{184440} (\bibinfo{year}{2005}).

\bibitem[{\citenamefont{{Christensen} et~al.}(2007)\citenamefont{{Christensen},
  {R{\o}nnow}, {McMorrow}, {Harrison}, {Perring}, {Enderle}, {Coldea},
  {Regnault}, and {Aeppli}}}]{chris1}
\bibinfo{author}{\bibfnamefont{N.~B.} \bibnamefont{{Christensen}}},
  \bibinfo{author}{\bibfnamefont{H.~M.} \bibnamefont{{R{\o}nnow}}},
  \bibinfo{author}{\bibfnamefont{D.~F.} \bibnamefont{{McMorrow}}},
  \bibinfo{author}{\bibfnamefont{A.}~\bibnamefont{{Harrison}}},
  \bibinfo{author}{\bibfnamefont{T.~G.} \bibnamefont{{Perring}}},
  \bibinfo{author}{\bibfnamefont{M.}~\bibnamefont{{Enderle}}},
  \bibinfo{author}{\bibfnamefont{R.}~\bibnamefont{{Coldea}}},
  \bibinfo{author}{\bibfnamefont{L.~P.} \bibnamefont{{Regnault}}},
  \bibnamefont{and} \bibinfo{author}{\bibfnamefont{G.}~\bibnamefont{{Aeppli}}},
  \bibinfo{journal}{Proceedings of the National Academy of Science}
  \textbf{\bibinfo{volume}{104}}, \bibinfo{pages}{15264}
  (\bibinfo{year}{2007}).

\bibitem[{\citenamefont{{Dalla Piazza} et~al.}(2015)\citenamefont{{Dalla
  Piazza}, {Mourigal}, {Christensen}, {Nilsen}, {Tregenna-Piggott}, {Perring},
  {Enderle}, {McMorrow}, {Ivanov}, and {R{\o}nnow}}}]{piazza}
\bibinfo{author}{\bibfnamefont{B.}~\bibnamefont{{Dalla Piazza}}},
  \bibinfo{author}{\bibfnamefont{M.}~\bibnamefont{{Mourigal}}},
  \bibinfo{author}{\bibfnamefont{N.~B.} \bibnamefont{{Christensen}}},
  \bibinfo{author}{\bibfnamefont{G.~J.} \bibnamefont{{Nilsen}}},
  \bibinfo{author}{\bibfnamefont{P.}~\bibnamefont{{Tregenna-Piggott}}},
  \bibinfo{author}{\bibfnamefont{T.~G.} \bibnamefont{{Perring}}},
  \bibinfo{author}{\bibfnamefont{M.}~\bibnamefont{{Enderle}}},
  \bibinfo{author}{\bibfnamefont{D.~F.} \bibnamefont{{McMorrow}}},
  \bibinfo{author}{\bibfnamefont{D.~A.} \bibnamefont{{Ivanov}}},
  \bibnamefont{and} \bibinfo{author}{\bibfnamefont{H.~M.}
  \bibnamefont{{R{\o}nnow}}}, \bibinfo{journal}{Nature Physics}
  \textbf{\bibinfo{volume}{11}}, \bibinfo{pages}{62} (\bibinfo{year}{2015}).

\bibitem[{\citenamefont{Chubukov et~al.}(1992)\citenamefont{Chubukov, Gagliano,
  and Balseiro}}]{ringsw1}
\bibinfo{author}{\bibfnamefont{A.}~\bibnamefont{Chubukov}},
  \bibinfo{author}{\bibfnamefont{E.}~\bibnamefont{Gagliano}}, \bibnamefont{and}
  \bibinfo{author}{\bibfnamefont{C.}~\bibnamefont{Balseiro}},
  \bibinfo{journal}{Phys. Rev. B} \textbf{\bibinfo{volume}{45}},
  \bibinfo{pages}{7889} (\bibinfo{year}{1992}).

\bibitem[{\citenamefont{Larsen et~al.}(2019)\citenamefont{Larsen, R\o{}mer,
  Janas, Treue, M\o{}nsted, Shaik, R\o{}nnow, and Lefmann}}]{ringnum1}
\bibinfo{author}{\bibfnamefont{C.~B.} \bibnamefont{Larsen}},
  \bibinfo{author}{\bibfnamefont{A.~T.} \bibnamefont{R\o{}mer}},
  \bibinfo{author}{\bibfnamefont{S.}~\bibnamefont{Janas}},
  \bibinfo{author}{\bibfnamefont{F.}~\bibnamefont{Treue}},
  \bibinfo{author}{\bibfnamefont{B.}~\bibnamefont{M\o{}nsted}},
  \bibinfo{author}{\bibfnamefont{N.~E.} \bibnamefont{Shaik}},
  \bibinfo{author}{\bibfnamefont{H.~M.} \bibnamefont{R\o{}nnow}},
  \bibnamefont{and} \bibinfo{author}{\bibfnamefont{K.}~\bibnamefont{Lefmann}},
  \bibinfo{journal}{Phys. Rev. B} \textbf{\bibinfo{volume}{99}},
  \bibinfo{pages}{054432} (\bibinfo{year}{2019}).

\bibitem[{\citenamefont{Martinelli et~al.}(2022)\citenamefont{Martinelli,
  Betto, Kummer, Arpaia, Braicovich, Di~Castro, Brookes, Moretti~Sala, and
  Ghiringhelli}}]{rixs6}
\bibinfo{author}{\bibfnamefont{L.}~\bibnamefont{Martinelli}},
  \bibinfo{author}{\bibfnamefont{D.}~\bibnamefont{Betto}},
  \bibinfo{author}{\bibfnamefont{K.}~\bibnamefont{Kummer}},
  \bibinfo{author}{\bibfnamefont{R.}~\bibnamefont{Arpaia}},
  \bibinfo{author}{\bibfnamefont{L.}~\bibnamefont{Braicovich}},
  \bibinfo{author}{\bibfnamefont{D.}~\bibnamefont{Di~Castro}},
  \bibinfo{author}{\bibfnamefont{N.~B.} \bibnamefont{Brookes}},
  \bibinfo{author}{\bibfnamefont{M.}~\bibnamefont{Moretti~Sala}},
  \bibnamefont{and}
  \bibinfo{author}{\bibfnamefont{G.}~\bibnamefont{Ghiringhelli}},
  \bibinfo{journal}{Phys. Rev. X} \textbf{\bibinfo{volume}{12}},
  \bibinfo{pages}{021041} (\bibinfo{year}{2022}).

\bibitem[{\citenamefont{Pekker and Varma}(2015)}]{higgs1}
\bibinfo{author}{\bibfnamefont{D.}~\bibnamefont{Pekker}} \bibnamefont{and}
  \bibinfo{author}{\bibfnamefont{C.}~\bibnamefont{Varma}},
  \bibinfo{journal}{Annual Review of Condensed Matter Physics}
  \textbf{\bibinfo{volume}{6}}, \bibinfo{pages}{269} (\bibinfo{year}{2015}).

\bibitem[{\citenamefont{Tohyama and Maekawa}(2000)}]{arpes}
\bibinfo{author}{\bibfnamefont{T.}~\bibnamefont{Tohyama}} \bibnamefont{and}
  \bibinfo{author}{\bibfnamefont{S.}~\bibnamefont{Maekawa}},
  \bibinfo{journal}{Superconductor Science and Technology}
  \textbf{\bibinfo{volume}{13}}, \bibinfo{pages}{R17} (\bibinfo{year}{2000}),
  \urlprefix\url{https://dx.doi.org/10.1088/0953-2048/13/4/201}.

\bibitem[{\citenamefont{Shastry and Shraiman}(1990)}]{ramansh}
\bibinfo{author}{\bibfnamefont{B.~S.} \bibnamefont{Shastry}} \bibnamefont{and}
  \bibinfo{author}{\bibfnamefont{B.~I.} \bibnamefont{Shraiman}},
  \bibinfo{journal}{Phys. Rev. Lett.} \textbf{\bibinfo{volume}{65}},
  \bibinfo{pages}{1068} (\bibinfo{year}{1990}).

\bibitem[{\citenamefont{Fleury and Loudon}(1968)}]{lframan}
\bibinfo{author}{\bibfnamefont{P.~A.} \bibnamefont{Fleury}} \bibnamefont{and}
  \bibinfo{author}{\bibfnamefont{R.}~\bibnamefont{Loudon}},
  \bibinfo{journal}{Phys. Rev.} \textbf{\bibinfo{volume}{166}},
  \bibinfo{pages}{514} (\bibinfo{year}{1968}).

\bibitem[{\citenamefont{Blumberg et~al.}(1996)\citenamefont{Blumberg,
  Abbamonte, Klein, Lee, Ginsberg, Miller, and Zibold}}]{ramanexp4}
\bibinfo{author}{\bibfnamefont{G.}~\bibnamefont{Blumberg}},
  \bibinfo{author}{\bibfnamefont{P.}~\bibnamefont{Abbamonte}},
  \bibinfo{author}{\bibfnamefont{M.~V.} \bibnamefont{Klein}},
  \bibinfo{author}{\bibfnamefont{W.~C.} \bibnamefont{Lee}},
  \bibinfo{author}{\bibfnamefont{D.~M.} \bibnamefont{Ginsberg}},
  \bibinfo{author}{\bibfnamefont{L.~L.} \bibnamefont{Miller}},
  \bibnamefont{and} \bibinfo{author}{\bibfnamefont{A.}~\bibnamefont{Zibold}},
  \bibinfo{journal}{Phys. Rev. B} \textbf{\bibinfo{volume}{53}},
  \bibinfo{pages}{R11930} (\bibinfo{year}{1996}).

\bibitem[{\citenamefont{Chubukov and Frenkel}(1995)}]{raman4}
\bibinfo{author}{\bibfnamefont{A.~V.} \bibnamefont{Chubukov}} \bibnamefont{and}
  \bibinfo{author}{\bibfnamefont{D.~M.} \bibnamefont{Frenkel}},
  \bibinfo{journal}{Phys. Rev. B} \textbf{\bibinfo{volume}{52}},
  \bibinfo{pages}{9760} (\bibinfo{year}{1995}).

\bibitem[{\citenamefont{Morr and Chubukov}(1997)}]{raman5}
\bibinfo{author}{\bibfnamefont{D.~K.} \bibnamefont{Morr}} \bibnamefont{and}
  \bibinfo{author}{\bibfnamefont{A.~V.} \bibnamefont{Chubukov}},
  \bibinfo{journal}{Phys. Rev. B} \textbf{\bibinfo{volume}{56}},
  \bibinfo{pages}{9134} (\bibinfo{year}{1997}).

\bibitem[{\citenamefont{Parkinson}(1969)}]{raman1}
\bibinfo{author}{\bibfnamefont{J.~B.} \bibnamefont{Parkinson}},
  \bibinfo{journal}{Journal of Physics C: Solid State Physics}
  \textbf{\bibinfo{volume}{2}}, \bibinfo{pages}{2012} (\bibinfo{year}{1969}).

\bibitem[{\citenamefont{Davies et~al.}(1971)\citenamefont{Davies, Chinn, and
  Zeiger}}]{raman2}
\bibinfo{author}{\bibfnamefont{R.~W.} \bibnamefont{Davies}},
  \bibinfo{author}{\bibfnamefont{S.~R.} \bibnamefont{Chinn}}, \bibnamefont{and}
  \bibinfo{author}{\bibfnamefont{H.~J.} \bibnamefont{Zeiger}},
  \bibinfo{journal}{Phys. Rev. B} \textbf{\bibinfo{volume}{4}},
  \bibinfo{pages}{992} (\bibinfo{year}{1971}).

\bibitem[{\citenamefont{Canali and Girvin}(1992)}]{raman3}
\bibinfo{author}{\bibfnamefont{C.~M.} \bibnamefont{Canali}} \bibnamefont{and}
  \bibinfo{author}{\bibfnamefont{S.~M.} \bibnamefont{Girvin}},
  \bibinfo{journal}{Phys. Rev. B} \textbf{\bibinfo{volume}{45}},
  \bibinfo{pages}{7127} (\bibinfo{year}{1992}).

\bibitem[{\citenamefont{Chelwani et~al.}(2018)\citenamefont{Chelwani, Baum,
  B\"ohm, Opel, Venturini, Tassini, Erb, Berger, Forr\'o, and
  Hackl}}]{ramanexp3}
\bibinfo{author}{\bibfnamefont{N.}~\bibnamefont{Chelwani}},
  \bibinfo{author}{\bibfnamefont{A.}~\bibnamefont{Baum}},
  \bibinfo{author}{\bibfnamefont{T.}~\bibnamefont{B\"ohm}},
  \bibinfo{author}{\bibfnamefont{M.}~\bibnamefont{Opel}},
  \bibinfo{author}{\bibfnamefont{F.}~\bibnamefont{Venturini}},
  \bibinfo{author}{\bibfnamefont{L.}~\bibnamefont{Tassini}},
  \bibinfo{author}{\bibfnamefont{A.}~\bibnamefont{Erb}},
  \bibinfo{author}{\bibfnamefont{H.}~\bibnamefont{Berger}},
  \bibinfo{author}{\bibfnamefont{L.}~\bibnamefont{Forr\'o}}, \bibnamefont{and}
  \bibinfo{author}{\bibfnamefont{R.}~\bibnamefont{Hackl}},
  \bibinfo{journal}{Phys. Rev. B} \textbf{\bibinfo{volume}{97}},
  \bibinfo{pages}{024407} (\bibinfo{year}{2018}).

\bibitem[{\citenamefont{Sandvik et~al.}(1998)\citenamefont{Sandvik, Capponi,
  Poilblanc, and Dagotto}}]{raman7}
\bibinfo{author}{\bibfnamefont{A.~W.} \bibnamefont{Sandvik}},
  \bibinfo{author}{\bibfnamefont{S.}~\bibnamefont{Capponi}},
  \bibinfo{author}{\bibfnamefont{D.}~\bibnamefont{Poilblanc}},
  \bibnamefont{and} \bibinfo{author}{\bibfnamefont{E.}~\bibnamefont{Dagotto}},
  \bibinfo{journal}{Phys. Rev. B} \textbf{\bibinfo{volume}{57}},
  \bibinfo{pages}{8478} (\bibinfo{year}{1998}).

\bibitem[{\citenamefont{Weidinger and Zwerger}(2015)}]{weidinger2015}
\bibinfo{author}{\bibfnamefont{S.~A.} \bibnamefont{Weidinger}}
  \bibnamefont{and} \bibinfo{author}{\bibfnamefont{W.}~\bibnamefont{Zwerger}},
  \bibinfo{journal}{The European Physical Journal B}
  \textbf{\bibinfo{volume}{88}}, \bibinfo{pages}{237} (\bibinfo{year}{2015}).

\bibitem[{\citenamefont{Ament et~al.}(2011)\citenamefont{Ament, van Veenendaal,
  Devereaux, Hill, and van~den Brink}}]{rixsrev}
\bibinfo{author}{\bibfnamefont{L.~J.~P.} \bibnamefont{Ament}},
  \bibinfo{author}{\bibfnamefont{M.}~\bibnamefont{van Veenendaal}},
  \bibinfo{author}{\bibfnamefont{T.~P.} \bibnamefont{Devereaux}},
  \bibinfo{author}{\bibfnamefont{J.~P.} \bibnamefont{Hill}}, \bibnamefont{and}
  \bibinfo{author}{\bibfnamefont{J.}~\bibnamefont{van~den Brink}},
  \bibinfo{journal}{Rev. Mod. Phys.} \textbf{\bibinfo{volume}{83}},
  \bibinfo{pages}{705} (\bibinfo{year}{2011}).

\bibitem[{\citenamefont{Betto et~al.}(2021)\citenamefont{Betto, Fumagalli,
  Martinelli, Rossi, Piombo, Yoshimi, Di~Castro, Di~Gennaro, Sambri, Bonn
  et~al.}}]{rixs1}
\bibinfo{author}{\bibfnamefont{D.}~\bibnamefont{Betto}},
  \bibinfo{author}{\bibfnamefont{R.}~\bibnamefont{Fumagalli}},
  \bibinfo{author}{\bibfnamefont{L.}~\bibnamefont{Martinelli}},
  \bibinfo{author}{\bibfnamefont{M.}~\bibnamefont{Rossi}},
  \bibinfo{author}{\bibfnamefont{R.}~\bibnamefont{Piombo}},
  \bibinfo{author}{\bibfnamefont{K.}~\bibnamefont{Yoshimi}},
  \bibinfo{author}{\bibfnamefont{D.}~\bibnamefont{Di~Castro}},
  \bibinfo{author}{\bibfnamefont{E.}~\bibnamefont{Di~Gennaro}},
  \bibinfo{author}{\bibfnamefont{A.}~\bibnamefont{Sambri}},
  \bibinfo{author}{\bibfnamefont{D.}~\bibnamefont{Bonn}}, \bibnamefont{et~al.},
  \bibinfo{journal}{Phys. Rev. B} \textbf{\bibinfo{volume}{103}},
  \bibinfo{pages}{L140409} (\bibinfo{year}{2021}).

\bibitem[{\citenamefont{Braicovich et~al.}(2010)\citenamefont{Braicovich,
  van~den Brink, Bisogni, Sala, Ament, Brookes, De~Luca, Salluzzo, Schmitt,
  Strocov et~al.}}]{rixs2}
\bibinfo{author}{\bibfnamefont{L.}~\bibnamefont{Braicovich}},
  \bibinfo{author}{\bibfnamefont{J.}~\bibnamefont{van~den Brink}},
  \bibinfo{author}{\bibfnamefont{V.}~\bibnamefont{Bisogni}},
  \bibinfo{author}{\bibfnamefont{M.~M.} \bibnamefont{Sala}},
  \bibinfo{author}{\bibfnamefont{L.~J.~P.} \bibnamefont{Ament}},
  \bibinfo{author}{\bibfnamefont{N.~B.} \bibnamefont{Brookes}},
  \bibinfo{author}{\bibfnamefont{G.~M.} \bibnamefont{De~Luca}},
  \bibinfo{author}{\bibfnamefont{M.}~\bibnamefont{Salluzzo}},
  \bibinfo{author}{\bibfnamefont{T.}~\bibnamefont{Schmitt}},
  \bibinfo{author}{\bibfnamefont{V.~N.} \bibnamefont{Strocov}},
  \bibnamefont{et~al.}, \bibinfo{journal}{Phys. Rev. Lett.}
  \textbf{\bibinfo{volume}{104}}, \bibinfo{pages}{077002}
  (\bibinfo{year}{2010}).

\bibitem[{\citenamefont{Minola et~al.}(2015)\citenamefont{Minola, Dellea,
  Gretarsson, Peng, Lu, Porras, Loew, Yakhou, Brookes, Huang et~al.}}]{rixs3}
\bibinfo{author}{\bibfnamefont{M.}~\bibnamefont{Minola}},
  \bibinfo{author}{\bibfnamefont{G.}~\bibnamefont{Dellea}},
  \bibinfo{author}{\bibfnamefont{H.}~\bibnamefont{Gretarsson}},
  \bibinfo{author}{\bibfnamefont{Y.~Y.} \bibnamefont{Peng}},
  \bibinfo{author}{\bibfnamefont{Y.}~\bibnamefont{Lu}},
  \bibinfo{author}{\bibfnamefont{J.}~\bibnamefont{Porras}},
  \bibinfo{author}{\bibfnamefont{T.}~\bibnamefont{Loew}},
  \bibinfo{author}{\bibfnamefont{F.}~\bibnamefont{Yakhou}},
  \bibinfo{author}{\bibfnamefont{N.~B.} \bibnamefont{Brookes}},
  \bibinfo{author}{\bibfnamefont{Y.~B.} \bibnamefont{Huang}},
  \bibnamefont{et~al.}, \bibinfo{journal}{Phys. Rev. Lett.}
  \textbf{\bibinfo{volume}{114}}, \bibinfo{pages}{217003}
  (\bibinfo{year}{2015}).

\bibitem[{\citenamefont{Peng et~al.}(2018)\citenamefont{Peng, Huang, Fumagalli,
  Minola, Wang, Sun, Ding, Kummer, Zhou, Brookes et~al.}}]{rixs4}
\bibinfo{author}{\bibfnamefont{Y.~Y.} \bibnamefont{Peng}},
  \bibinfo{author}{\bibfnamefont{E.~W.} \bibnamefont{Huang}},
  \bibinfo{author}{\bibfnamefont{R.}~\bibnamefont{Fumagalli}},
  \bibinfo{author}{\bibfnamefont{M.}~\bibnamefont{Minola}},
  \bibinfo{author}{\bibfnamefont{Y.}~\bibnamefont{Wang}},
  \bibinfo{author}{\bibfnamefont{X.}~\bibnamefont{Sun}},
  \bibinfo{author}{\bibfnamefont{Y.}~\bibnamefont{Ding}},
  \bibinfo{author}{\bibfnamefont{K.}~\bibnamefont{Kummer}},
  \bibinfo{author}{\bibfnamefont{X.~J.} \bibnamefont{Zhou}},
  \bibinfo{author}{\bibfnamefont{N.~B.} \bibnamefont{Brookes}},
  \bibnamefont{et~al.}, \bibinfo{journal}{Phys. Rev. B}
  \textbf{\bibinfo{volume}{98}}, \bibinfo{pages}{144507}
  (\bibinfo{year}{2018}).

\bibitem[{\citenamefont{Dean et~al.}(2012)\citenamefont{Dean, Springell,
  Monney, Zhou, Pereiro, Bo{\v{z}}ovi{\'{c}}, Dalla~Piazza, R{\o}nnow,
  Morenzoni, van~den Brink et~al.}}]{rixs5}
\bibinfo{author}{\bibfnamefont{M.~P.~M.} \bibnamefont{Dean}},
  \bibinfo{author}{\bibfnamefont{R.~S.} \bibnamefont{Springell}},
  \bibinfo{author}{\bibfnamefont{C.}~\bibnamefont{Monney}},
  \bibinfo{author}{\bibfnamefont{K.~J.} \bibnamefont{Zhou}},
  \bibinfo{author}{\bibfnamefont{J.}~\bibnamefont{Pereiro}},
  \bibinfo{author}{\bibfnamefont{I.}~\bibnamefont{Bo{\v{z}}ovi{\'{c}}}},
  \bibinfo{author}{\bibfnamefont{B.}~\bibnamefont{Dalla~Piazza}},
  \bibinfo{author}{\bibfnamefont{H.~M.} \bibnamefont{R{\o}nnow}},
  \bibinfo{author}{\bibfnamefont{E.}~\bibnamefont{Morenzoni}},
  \bibinfo{author}{\bibfnamefont{J.}~\bibnamefont{van~den Brink}},
  \bibnamefont{et~al.}, \bibinfo{journal}{Nature Materials}
  \textbf{\bibinfo{volume}{11}}, \bibinfo{pages}{850} (\bibinfo{year}{2012}).

\bibitem[{\citenamefont{Le~Tacon et~al.}(2011)\citenamefont{Le~Tacon,
  Ghiringhelli, Chaloupka, Sala, Hinkov, Haverkort, Minola, Bakr, Zhou,
  Blanco-Canosa et~al.}}]{Tacon}
\bibinfo{author}{\bibfnamefont{M.}~\bibnamefont{Le~Tacon}},
  \bibinfo{author}{\bibfnamefont{G.}~\bibnamefont{Ghiringhelli}},
  \bibinfo{author}{\bibfnamefont{J.}~\bibnamefont{Chaloupka}},
  \bibinfo{author}{\bibfnamefont{M.~M.} \bibnamefont{Sala}},
  \bibinfo{author}{\bibfnamefont{V.}~\bibnamefont{Hinkov}},
  \bibinfo{author}{\bibfnamefont{M.~W.} \bibnamefont{Haverkort}},
  \bibinfo{author}{\bibfnamefont{M.}~\bibnamefont{Minola}},
  \bibinfo{author}{\bibfnamefont{M.}~\bibnamefont{Bakr}},
  \bibinfo{author}{\bibfnamefont{K.~J.} \bibnamefont{Zhou}},
  \bibinfo{author}{\bibfnamefont{S.}~\bibnamefont{Blanco-Canosa}},
  \bibnamefont{et~al.}, \bibinfo{journal}{Nature Physics}
  \textbf{\bibinfo{volume}{7}}, \bibinfo{pages}{725 } (\bibinfo{year}{2011}),
  \bibinfo{note}{and references therein}.

\bibitem[{\citenamefont{Suzuura et~al.}(1996)\citenamefont{Suzuura, Yasuhara,
  Furusaki, Nagaosa, and Tokura}}]{ir1}
\bibinfo{author}{\bibfnamefont{H.}~\bibnamefont{Suzuura}},
  \bibinfo{author}{\bibfnamefont{H.}~\bibnamefont{Yasuhara}},
  \bibinfo{author}{\bibfnamefont{A.}~\bibnamefont{Furusaki}},
  \bibinfo{author}{\bibfnamefont{N.}~\bibnamefont{Nagaosa}}, \bibnamefont{and}
  \bibinfo{author}{\bibfnamefont{Y.}~\bibnamefont{Tokura}},
  \bibinfo{journal}{Phys. Rev. Lett.} \textbf{\bibinfo{volume}{76}},
  \bibinfo{pages}{2579} (\bibinfo{year}{1996}).

\bibitem[{\citenamefont{Lorenzana and Eder}(1997)}]{ir4}
\bibinfo{author}{\bibfnamefont{J.}~\bibnamefont{Lorenzana}} \bibnamefont{and}
  \bibinfo{author}{\bibfnamefont{R.}~\bibnamefont{Eder}},
  \bibinfo{journal}{Phys. Rev. B} \textbf{\bibinfo{volume}{55}},
  \bibinfo{pages}{R3358} (\bibinfo{year}{1997}).

\bibitem[{\citenamefont{Lorenzana et~al.}(1999)\citenamefont{Lorenzana, Eroles,
  and Sorella}}]{ir5}
\bibinfo{author}{\bibfnamefont{J.}~\bibnamefont{Lorenzana}},
  \bibinfo{author}{\bibfnamefont{J.}~\bibnamefont{Eroles}}, \bibnamefont{and}
  \bibinfo{author}{\bibfnamefont{S.}~\bibnamefont{Sorella}},
  \bibinfo{journal}{Phys. Rev. Lett.} \textbf{\bibinfo{volume}{83}},
  \bibinfo{pages}{5122} (\bibinfo{year}{1999}).

\end{thebibliography}

\end{document}